\documentclass[twocolumn,showpacs,amsmath,preprintnumbers,amssymb,10pt]{revtex4}
\usepackage{graphicx}
\usepackage{color}
\begin{document}

\title{Dynamical self-trapping of two-dimensional binary solitons in cross-combined linear and nonlinear optical lattices}
\author{K.K. Ismailov $^{1,3}$, G.A. Sekh $^2$, and Mario Salerno $^3$ }
\affiliation{
$^1$ Physical-Technical Institute, Uzbek Academy of Sciences, 100084, Tashkent, Uzbekistan \\
$^2$ Department of Physics, Kazi Nazrul University, Assansol-713340, India \\
$^3$ Dipartimento di Fisica ``E.R. Caianiello", and INFN Gruppo Collegato di Salerno,  Universit\'a di Salerno, Via Giovanni Paolo II, 84084 Fisciano, Salerno, Italy}
\date{\today}

\begin{abstract}
Dynamical and self-trapping properties of two-dimensional (2D) binary mixtures of Bose-Einstein condensates (BECs)  in  
cross-combined lattices consisting of a one-dimensional (1D) linear optical lattice (LOL) in the x-direction for the 
first component and a 1D non linear optical lattice (NOL) in the $y$-direction for the second component, are analytically and 
numerically investigated. The existence and stability of 2D binary matter wave solitons in these settings is 
demonstrated both by variational analysis and by direct numerical integration of the coupled Gross-Pitaevskii  
equations (GPE). We find that in absence of the NOL binary solitons, stabilized by the action of the 1D 
LOL and by the attractive inter-component interaction can freely move in the $y-$direction. In the presence of the 
NOL we find, quite remarkably, the existence of threshold curves in the parameter space separating regions where 
solitons can  move, from regions where the solitons  become dynamically self-trapped. The mechanism underlying  
the dynamical self-trapping phenomenon (DSTP) is qualitatively understood in terms of a dynamical barrier induced by the 
the NOL similar to the  Peirls-Nabarro barrier of solitons in discrete lattices. DSTP is numerically demonstrated  for binary 
solitons that are put in motion  both by phase imprinting and  by the action of external potentials applied in the $y-$direction. 
In the latter case we show that the trapping action of the NOL allows maintaining a 2D binary soliton at rest in a non-equilibrium 
position of a parabolic trap, or to prevent it from falling under the action of gravity. Possible applications of the results are also briefly discussed. 
\end{abstract}

\pacs{67.85.Hj, 03.75.Lm, 03.75.Kk, 67.85.Jk}
\maketitle
% % % % % % % % % % % % % % % % % % % % % % % % % %%%%%%%%%%%%%%%%%%%%%%%%%%%%%%%%%%%%%%%%%%%%%%%%%%%%%%%%%%%%%%%%%%%%%%%%%%%%%%%%%%%%%%%%%%%%
\section{Introduction}
Bose-Einstein condensates (BECs) of ultracold atoms trapped in optical lattices (OLs) are considered as ideal systems for 
realizing and understanding various phenomena of condensed matter and nonlinear physics. Experimental flexibility 
of controlling the system parameters over a wide range has  made it possible to observe phenomena like 
Bloch oscillations, dynamic localization, Landau-Zener tunneling, superfluid-Mott transition occurring in linear 
optical lattices (LOL)~\cite{morsch, dahan, boris, sekh, wilkinson, pathick, Li1, bloch, konotop}.  In a mean field 
description of the condensate, the nonlinearity that arises from the interatomic interactions if suitably balanced 
by dispersion, allows the formation of matter waves solitons~\cite{gap,gap1, Chai}. This is particularly true 
for one dimensional settings where solitons are very  stable and quite generic in the whole parameter space. 

In higher dimensions and in the presence of  attractive interactions, the occurrence of delocalization transitions~\cite{BS04} 
and the appearance of collapse phenomena induce criticality on the existence of stable solitons~\cite{Schmied, Katsimiga,Li}. The 
possibility of their stabilization by means of periodic potentials  was  demonstrated 
in~\cite{salerno1, salerno-PRA04}. In particular, in~\cite{salerno-PRA04} it was shown  that periodic  potentials of co-dimension 1 (i.e. potentials whose dimension is that of the  full space minus one) can support stable solitons both in 2D and in 3D attractive BECs. 

In addition to LOLs, the efficiency in controlling nonlinear interactions in time and space by means of magnetic or 
optically induced Feshbach resonances, has allowed the introduction of the so-called nonlinear optical lattice 
(NOL), i.e. a lattice induced by space dependent interatomic interactions. The effective potential produced by a NOL 
can be periodic or localized depending on whether the density of matter is periodic or localized. 

In one dimensional  settings NOLs have been shown to be very useful  to eliminate destructive dynamical instabilities, 
such as those arising in Bloch oscillations of gap-solitons moving in accelerated LOL~\cite{salerno}. 
In the multidimensional case it is proven  that 2D localized BEC can be stable in 1D cross-combined linear and nonlinear 
optical lattices~\citep{Tomio}, but neither a 1D NOL  nor a 2D NOL are sufficient to hold stable 
2D BECs~\cite{cite4}. In nonlinear optics contexts it was recently shown that NOLs of the form of arrays of Kerr-
nonlinear cylinders embedded into linear~\cite{ref17} or saturable host media can support stable 2D 
solitons~\cite{ref18}. We also remark that the  existence of 2D Bose-Einstein condensates has been  realized experimentally 
in a combination of harmonic and lattice potentials~\cite{2DBEC,2DBEC1,burger} and all the studies reported above 
refer to the case of ordinary, i.e. 
single component, BECs.

Multi-dimensional solitons of  binary  BEC mixtures are more involved and much less investigated~\cite{ma}. In 
particular, we mention the results in~\cite{ref19} showing that either a 2D LOL or a 2D NOL applied only to one 
component, is sufficient to stabilize binary 2D BEC solitons against collapse or decay. In the case of a LOL the 
binary solitons were shown to be stable almost on all range of their existence, while in the NOL case there 
were restrictions on the number of atoms in the component affected by the lattice. In all these cases, the applied 
OLs (either linear or nonlinear) were two-dimensional and the resulting excitation intrinsically localized.

On the other hand, it is interesting to investigate 2D solitons of  binary BEC mixtures trapped in lower 
dimensional OLs of different nature. In this respect we remark that, in contrast with 1D LOLs, a 1D NOL is unable 
to sustain stable 2D solitons and therefore in a cross-combined  setting  (i.e. a 1D  LOL and a 1D NOL acting in 
different directions) binary solitons can exist only if the attractive inter-component interaction is sufficiently strong. Moreover, the different nature of the two lattices could lead to novel dynamical phenomena.

The aim of the paper is to investigate properties  of 2D solitons of binary BEC mixtures trapped in cross-combined  
OLs  consisting  of  a 1D LOL acting on one component along the x-direction and a 1D NOL acting on the other 
component along the $y$-direction. For this we use the  variational analysis (VA),  direct numerical integrations of the 
coupled  Gross-Pitaevskii equations (GPE) and the Vakhitov-Kolokolov (VK) criterion \cite{kolo} to investigate the existence and 
the stability of 2D binary solitons in such cross-combined OLs. We show  that in absence of the NOL, binary 2D solitons, stabilized 
by the action of the 1D  LOL and by the attractive inter-component interaction, can be put in action by phase imprinting and to  
move freely in the $y-$direction. In the presence of the  NOL, quite remarkably, we find the existence of threshold curves in the 
parameter space separating regions where solitons can  move from regions where they becomes dynamically self-trapped by the NOL. 

The mechanism underlying the dynamical self-trapping phenomenon (DSTP) is qualitatively explained  in terms of a dynamical barrier induced by the the NOL that resemble  the  Peirls-Nabarro barrier~\cite{kivshar-campbell} of solitons in discrete lattices. The DSTP is demonstrated  for binary solitons that are put in motion  both by phase imprinting and by the 
action of external potentials applied in the $y-$direction such as a parabolic trap or a linear ramp potential. In these cases 
we show that the DSTP allows to hold a  2D binary soliton at rest in a non-equilibrium position of a parabolic trap and  to  
prevent the soliton from falling under the action of the gravity. 

The paper is organized as follows. In section II, we introduce the model equations and in sections III we use the variational analysis and numerical GPE integrations  to investigate existence and stability of 2D binary isotropic and anisotropic  
solitons. In Section IV  we use the phase imprinting method to put stationary solitons in motion and to show the occurrence of the  the DSTP. In Sec. V the occurrence of the DSTP in external parabolic traps and linear ramp potentials is demonstrated. In  Sec. VI the possible relevance  of the DSTP  for application is briefly discussed and the main results are summarized.

%%%%%%%%%%%%%%%%%%%%%%%%%%%%%%%%%%%%%%%%%%%%%%%%%%%%%%%%%%%%%
\section{Model equations. }
%%%%%%%%%%%%%%%%%%%%%%%%%%%%%%%%%%%%%%%%%%%%%%%%%%%%%%%%%%%%%

At absolute zero temperature the ground state wavefunctions of a 2D binary BEC mixture can be described in the mean field  approximation by the following coupled Gross-Pitaevskii (GP) equations:
\begin{eqnarray}
\label{eq1}
&&i\frac{\partial\psi}{\partial t}=-\left(\nabla^2-V_L-\gamma_1 |\psi |^2- \gamma _{12} |\phi |^2\right) \psi, \\
&&i\frac{\partial\phi}{\partial t}\!=\!-\!\left(\nabla^2\!-\!V_{NL}|\phi |^2\!-\!\gamma_2 |\phi |^2\!-\! \gamma 
_{12}|\psi|^2\right)\!\phi,
\label{eq2}
\end{eqnarray}
where $\nabla^2$ denotes the 2D Laplacian and  $V_{L}$, $V_{NL}$ are periodic real functions modeling a LOL in the $x-$direction and a NOL in the $y-$direction, respectively, of the form:
\begin{equation}
V_{L}=V_1 \cos(2 x),\;\;\;\;\;\;  V_{NL}=V_2 \cos(2 y).
\label{OLNOL}
\end{equation}
In Eqs.~(\ref{eq1}),~(\ref{eq2}), $\psi$ and $\phi$ represent the component wavefunctions while 
the nonlinear coefficients $\gamma_i$, $i=1, 2, 12$ stand for $3D$ coupling constants  corresponding to the $s$-wave scattering 
lengths $a^{(i)}_{s}$, i.e. $\gamma_i=4\pi \hbar^2 a^{(i)}_{s}/m$, with  $m$, $l_z=\sqrt{\hbar/(m \omega_z)}$, $\omega_z$, denoting 
the atom mass, the  transverse oscillator length and the transverse frequency, respectively. The above  GPEs are written in 
dimensionless units obtained by replacing $t$ by $(\hbar/E_r)\,t$, $r\equiv (x,y)$ by $r/k$, with  $E_r=\hbar^2 k^2/(2 m)$ the 
recoil energy of the lattices, and $V_j$, $\gamma_j$ are measured in the units of $E_r$ and $E_r/k^2$, respectively.

It is also worth to note from Eqs.~(\ref{eq1}),~(\ref{eq2}), that while the first component is trapped by the potential 
$V_{L}$  acting in the x-direction, the second component is subjected to a nonlinear optical lattice $V_{NL}$ acting in the y-
direction. For this we assume the BEC components be associated to two hyperfine levels that are far detuned so that the laser used 
for the trapping of one component can be considered negligible for the other component and viceversa. 
The spatial modulation of the inter-atomic scattering length, can be produced by the optically induced Feshbach resonance 
technique, with the background scattering length assumed detuned to zero with appropriate experimental conditions. Moreover, the 
harmonic trap used to create the condensate is assumed to be weak enough to affect matter waves localized in the central part of 
the trap~\cite{konotop}.
%%%%%%%%%%%%%%%%%%%%%%%%%%%%%%%%%%%%%%%%%%%%%%%%%%%%%%%%%%%%%%%%%%%%%%%%%%%%%%%%%%%%
%
\begin{figure*}
\hskip -0.6cm
\centerline{\includegraphics[scale=0.45]{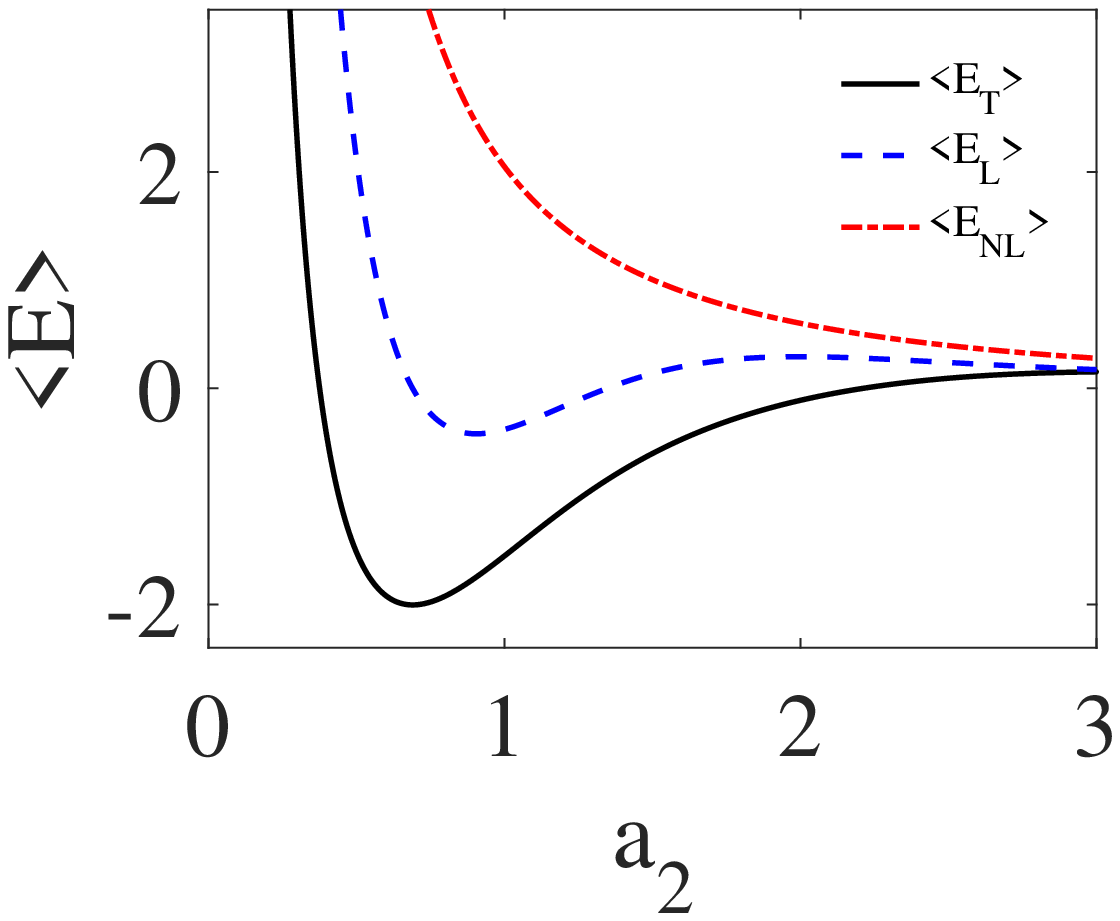}
\includegraphics[trim=0cm 0cm 1cm 0cm, clip=true,scale=0.45]{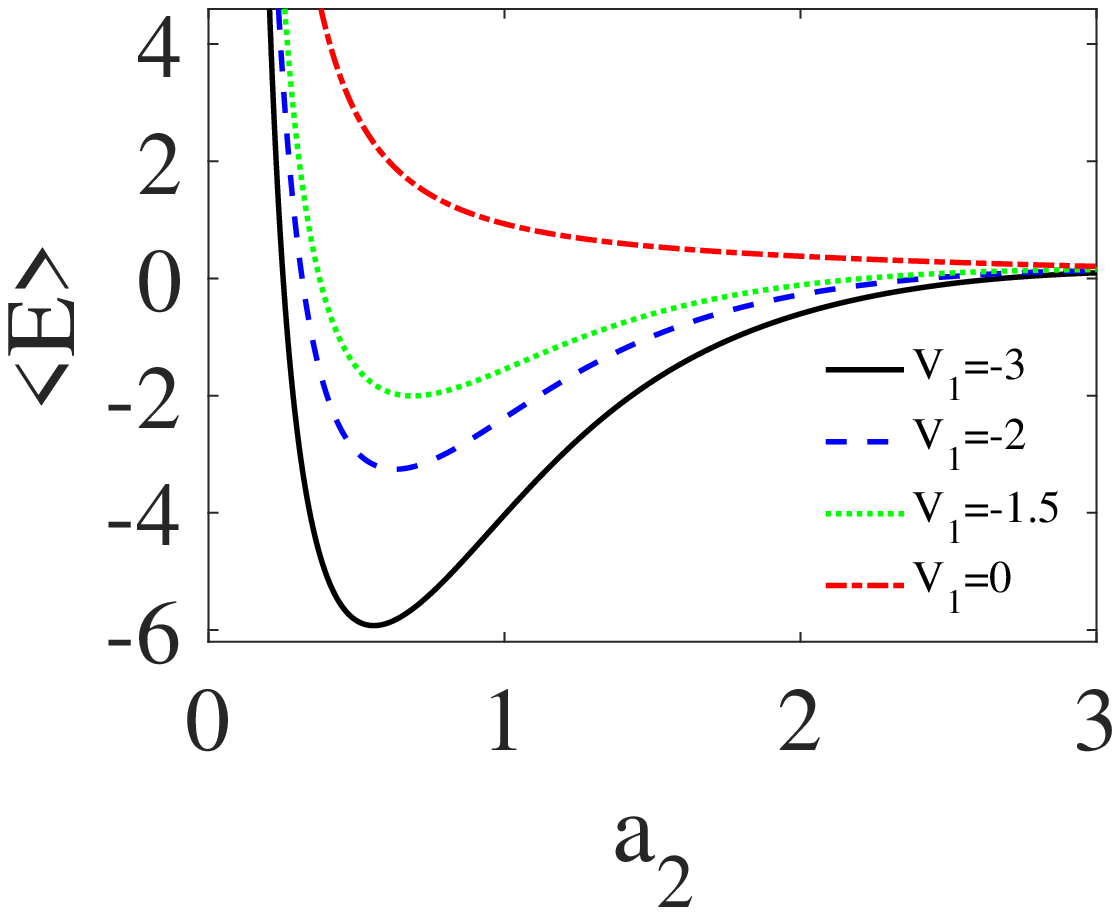}
\includegraphics[trim=0cm 0cm 1cm 0cm, clip=true,scale=0.45]{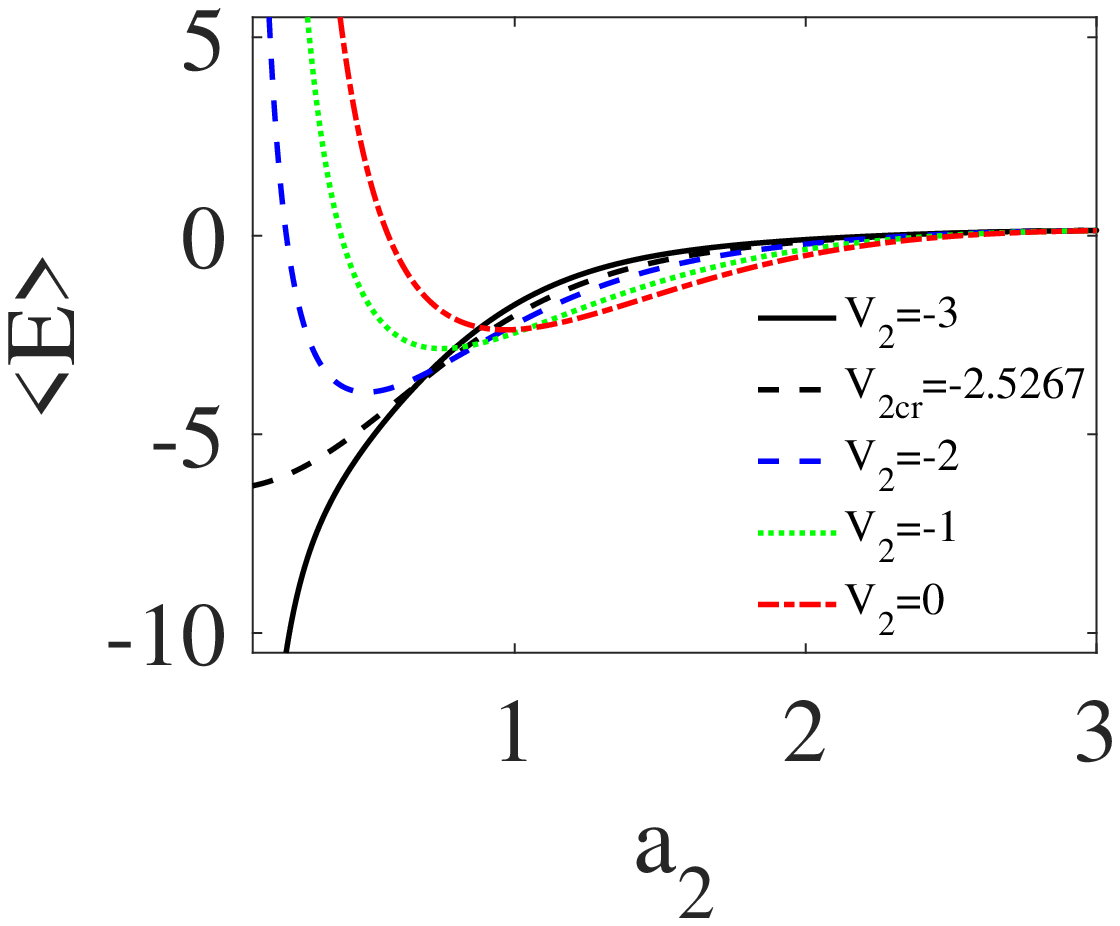}}
\caption{\small
{\bf Left panel}. Effective  energy-potential $\langle E\rangle$ versus  $a_2$ for uncoupled symmetric ($a_i=b_i$) 
components (blue dashed and red dash-dotted curves), and for the coupled  $\gamma_{12}=-2$ case  (solid black 
curve).  Other parameters are fixed as: $\gamma_{1}=-2$, $\gamma_{2}=0$, $V_1=-1.5$, $V_2=-1.5$,  ${\cal N}_1=3.5$ 
and ${\cal N}_2=2.5$. {\bf Middle panel}. Same as in the left panel but for $\gamma_{12}=-2$, $V_2=-1.5$ 
and  different values of $V_1$ indicated in the figure. Other parameters are fixed as in the left panel. Values of 
$a_1$ and $a_2$  are calculated  at the minima of the depicted bonding curves. {\bf Right panel}. Same as in the 
middle  panel but for $V_1=-2$ and different values of $V_2$ indicated in the figure. Other parameters are fixed as 
in the middle panel.
}
\label{fig1}
\end{figure*}
%
%%%%%%%%%%%%%%%%%%%%%%%%%%%%%%%%%%%%%%%%%%%%%%%%%%%%%%%%%%%%%%%%%%%%%%%%%%%%%%%%%%

\section{2D binary solitons: VA  and numerical results}
In this section we investigate  existence and stability properties of 2D binary solitons by means of the VA~\cite{anderson} based 
on the Gross-Pitaevskii energy density:
\begin{eqnarray}
E[\psi,\phi]&=&|\nabla \psi|^2+|\nabla \phi|^2+ V_1 \cos(2  x) |\psi|^2+\frac{1}
{2}\gamma_1|\psi|^4\nonumber \\
&+&\frac{1}{2}V_2 \cos(2 y)|\phi|^4+\frac{1}{2}\gamma_2|\phi|^4+\gamma_{12} 
|\phi|^2 |\psi|^2.
\label{eq3}
\end{eqnarray}
Results are then  comparison with direct numerical integration of the GPE system. With respect to perturbation theory, the VA 
represents  a simple  effective method to get properties of the ground state wavefunction . The efficiency and accuracy of the 
method depend in large part on the  choice of the trial function that should reflect from one side the properties of the system 
(symmetries, norms, etc), and from the other side 
should be simple enough to allow an analytical evaluation of the energy. The trapping potentials, the type of solutions searched 
(localized, extended) and the parameter region in which they are searched also play an important role for the choice. Thus, for 
example, looking for localized binary  matter-waves in the limit of negligible trapping 
potentials one could take the trial function as a  product of 1D solitons (sech-sech trial functions), while in the 
limit of negligible nonlinearities the Gaussian ansatz may be more appropriate. We tried  both ansatz  for our setting and we 
found that the Gaussian ansatz allows an analytical expression of the energy for generic values of the parameters (see below), 
while with the sech-sech ansatz this is possible only when the NOL is absent~\cite{note3}.  In the following, therefore, we assume 
a Gaussian  ansatz for the component wavefunctions:
\begin{eqnarray}
\psi(x,y)=A_1 \,\exp[-x^2/(2 a_1^2)-y^2
/(2 b_1^2)],
\label{eq4}
\end{eqnarray}
\begin{eqnarray}
\phi(x,y)=A_2\,\exp[-x^2/(2 a_2^2)-y^2/(2 b_2^2)],
\label{eq5}
\end{eqnarray}
with corresponding norms given by: 
\begin{eqnarray}
{\cal N}_1&=&\int|\psi|^2 dx dy = \pi a_1 b_1 A_1^2 \,, \nonumber \\
{\cal N}_2&=&\int|\phi|^2 dx dy = \pi a_2 b_2 A_2^2 \,.  
\label{norm}
\end{eqnarray}
%
%%%%%%%%%%%%%%%%%%%%%%%%%%%%%%%%%%%%%%%%%%%%%%%%%%%%%%%%%%%%%%%%%%%%%%%%%%%%%%%%%
%
\begin{figure*}
\hskip -1cm
\centerline{\includegraphics[scale=0.45]{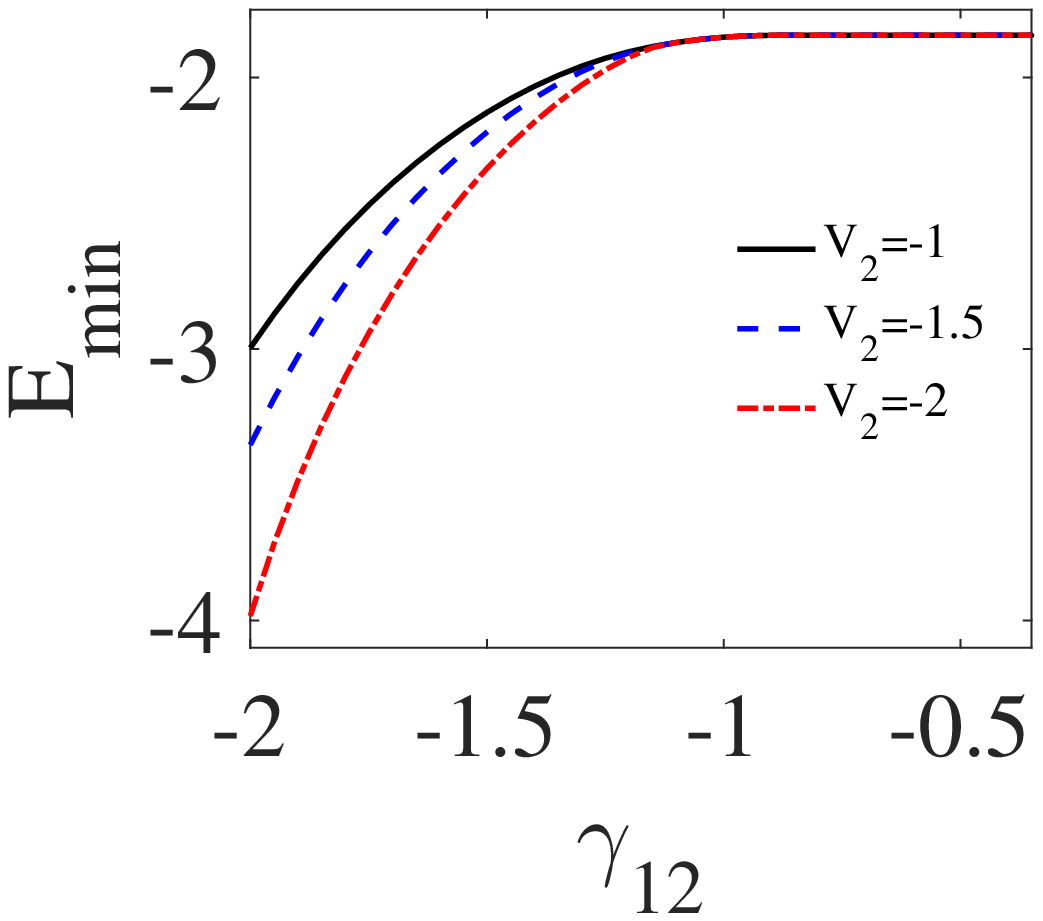}
\includegraphics[trim=0cm 0cm 1cm 0cm, clip=true,scale=0.45]{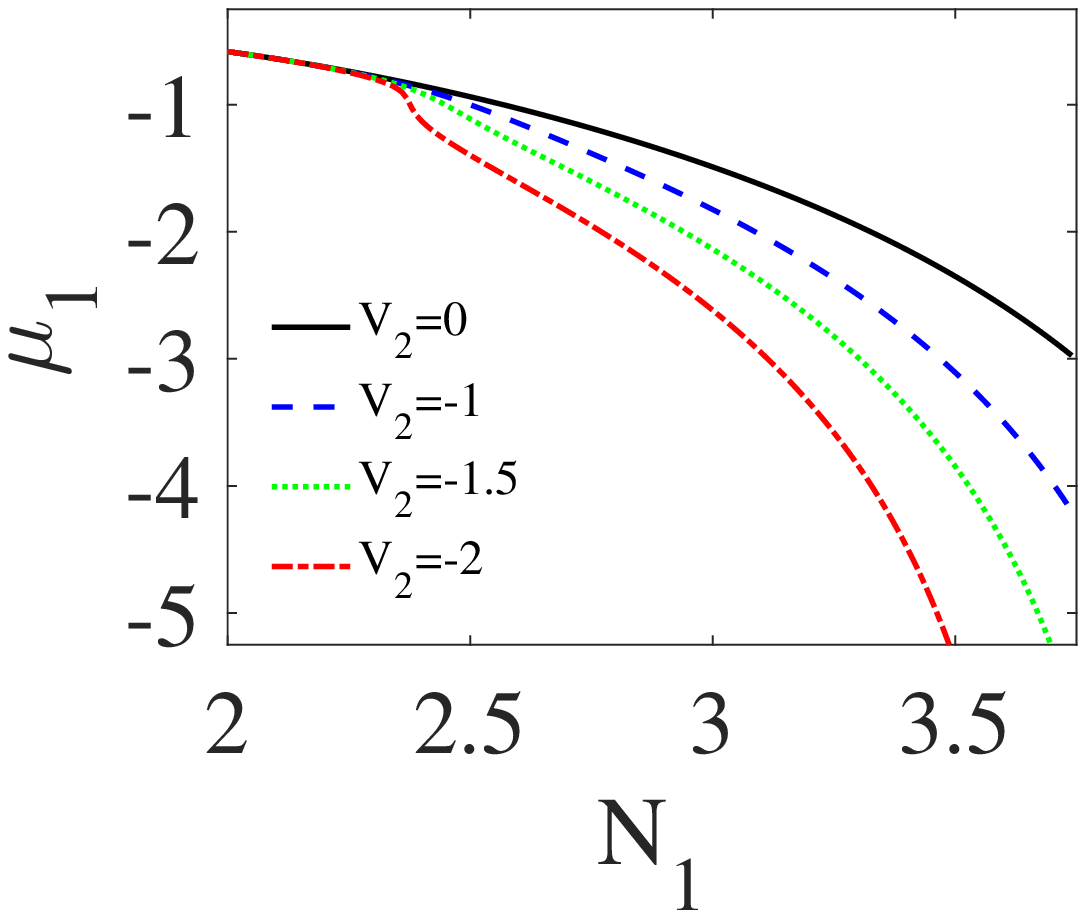}
\includegraphics[trim=0cm 0cm 1cm 0cm, clip=true,scale=0.45]{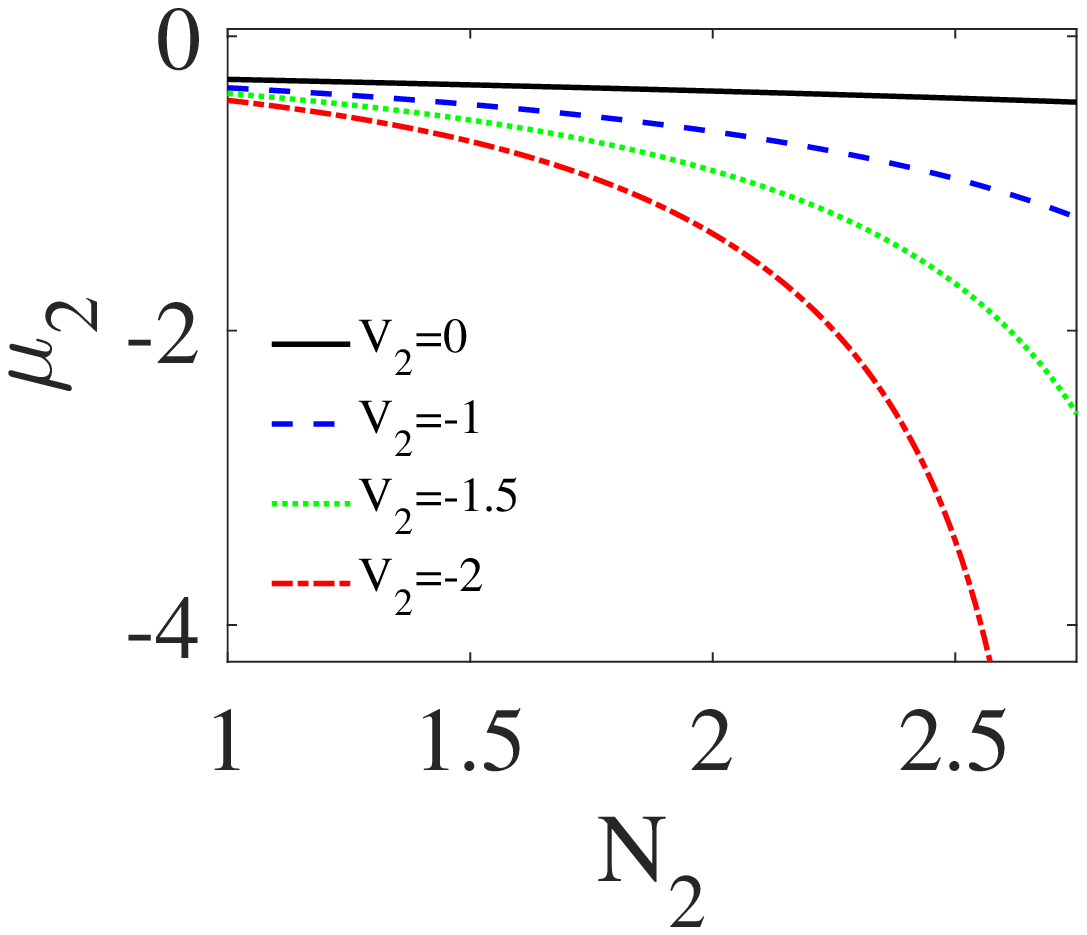}}
\hskip -1cm
\caption{\small
{\bf Left panel}. Minimum energy ($E_{min}$) as a function of  $\gamma_{12}$ for $V_2=-1$ (solid black), $-1.5$ 
(dashed blue) and $-2$ (dash-dotted red). Other parameters of the system are fixed as: $\gamma_{1}=-2$, $V_1=-2$, 
${\cal N}_1=3.5$, and ${\cal N}_2=2.5$.
{\bf  Middle and right panels}. Chemical potentials, $\mu_i$, versus number of atoms, $N_i$, of  first (middle) and 
second (right) component, for strengths ($V_2$) of the NOL indicated in the figure. Other parameters are fixed as 
$\gamma_{1}=-2$, $\gamma_{12}=-2.0$, $V_1=-2.0$.}
\label{fig2}
\end{figure*}
%
%%%%%%%%%%%%%%%%%%%%%%%%%%%%%%%%%%%%%%%%%%%%%%%%%%%%%%%%%%%%%%%%%%%%%%%%%%%%%
%
Here $a_j$, $b_j$, $j=1,2$ denote the widths of the two component profiles in the $x$ and $y$ directions, respectively, and  $A_j$ 
the corresponding profile amplitudes. Similar trial solutions were used in~\cite{salerno1,salerno-PRA04,Tomio,cite4} to describe 
solitons ($a_j/\pi \leq 1$) in single-component two-dimensional BECs in OLs. The integration of the energy density (\ref{eq3}) on 
the whole  $x-y$ plane leads to the following effective energy:
\begin{eqnarray}
&& \langle E\rangle=\sum_{j=1}^{2}\left[\frac{{\cal N}_j}{2} \left(\frac{1}{a_j^2}+\frac{1}
{b_j^2}\right)+\frac{\gamma_j {\cal N}_j^2}{4\pi a_j b_j}\right]+V_1 {\cal N}_1 e^{-a_1^2} \nonumber\\
&&
\;\;\;\;\;\;\;\; + \;\frac{V_2{\cal N}_2^2}{4\pi a_2 b_2} e^{-\frac{b_2^2}2}+
\frac{\gamma_{12}{\cal N}_1{\cal N}_2}{\pi[(a_1^2+a_2^2)(b_1^2+b_2^2)]^{\frac12}}.
\label{eq6}
\end{eqnarray}
Notice that in writing this equation we used Eqs.~(\ref{norm}) to eliminate $A_j$ in favor of  $[{\cal N}_j/(\pi a_j 
b_j)]^{\frac12}$.
The stationary conditions $\frac{\partial \langle E \rangle}{\partial a_i}=0, \frac{\partial \langle E \rangle}
{\partial b_i}=0,\, i=1,2$   of the energy function in Eq. (\ref{eq6}) provide a system of four equations, namely,
\begin{eqnarray}
\label{eq7a}
&& \frac{1}{a_1^3}+\frac{\gamma_1 {\cal N}_1}{4 \pi a_1^2 b_1}+\frac{a_1}{\pi} \gamma_{12} {\cal N}_2 F_1 + 2 V_1 
a_1 e^{-a_1^2} =0, \\
&& \frac{1}{b_1^3}+\frac{{\gamma_1} {{\cal N}_1}}{4 {a_1} {b_1}^2 \pi}+ \frac{b_1}{\pi} \gamma_{12} {\cal N}_2 
F_2=0,
\label{eq9a}
\end{eqnarray}
\begin{equation}
\label{eq8a}
\frac{1}{a_2^3}+\frac{a_2}{\pi} \gamma_{12} {\cal N}_1 F_1
+{\cal N}_2\frac{\gamma_2 + e^{-\frac{b_2^2}{2}} V_2}{4\pi a_2^2 b_2}=0,
\end{equation}
\begin{equation}
\frac{1}{b_2^3}+
{\cal N}_2\frac{\gamma_2+(1+{b_2}^2) e^{-\frac{{b_2}^2}2}{V_2}}{4\pi a_2{b_2}^2}
+\frac{b_2}{\pi} \gamma_{12} {\cal N}_1 F_2 =0,
\label{eq10a}
\end{equation}
with $F_1=({a_1}^2+{a_2}^2)^{-3/2} (b_1^2+b_2^2)^{-1/2}$ and $F_2$ obtained from $F_1$ by interchanging the $b$'s 
with $a$'s. To obtain parameters for existence of binary BEC solitons  in general one  must solve Eqs.~(\ref{eq7a})-
(\ref{eq10a}) numerically. For some specific choice of the system parameters, the energy of the system displays a 
minimum that is in general  negative. Binary solitons follow from Eqs. (\ref{eq4}), (\ref{eq5}) with parameters $A_i, a_i, b_i$, 
$i=1,2$  determined in correspondence of the energy minimum. 

The existence of a minimum, however, does not guarantee the stability of the soliton, which should then be specifically 
investigated. This can be done analytically within a VA approach by means of the VK criterion~\cite{vkc, kolo} according to which a 
binary soliton is stable if the change of a corresponding conserved (numbers of atoms ${\cal N}_j$) quantity with respect to its 
conjugated variable (chemical potential $\mu_j$) is negative. 

In this respect it is worth to  note that for stationary solutions 
\begin{equation}
\phi \equiv\phi(x,y) \,\exp(-i \mu_{1} t), \;\;\;
\psi \equiv \psi(x,y) \,\exp(-i \mu_{2} t),   \nonumber   
\end{equation}
the GPEs in~(\ref{eq1}),~(\ref{eq2}) can be rewritten as
\begin{eqnarray}
\label{eq11}
\mu_1=\frac{1}{{\cal N}_1}\int(|\nabla \psi|^2+V_L|\psi|^2&+&\gamma_1 |\psi|^4\nonumber\\
&+&\gamma _{12} |\phi |^2 |\psi|^2) d\tau, \\
\mu_{2}=\frac{1}{{\cal N}_2}\int(|\nabla \phi|^2+V_{NL}|\phi |^4&+&\gamma_{2}|\phi |^4\nonumber\\
&+&\gamma _{12} |\psi |^2 |\phi|^2) d\tau,
\label{eq12}
\end{eqnarray}
with $d\tau=dxdy$. Substituting  Eqs.~(\ref{OLNOL}),~(\ref{eq4}),~(\ref{eq5}) into Eqs.~(\ref{eq11}),~(\ref{eq12}) 
we get:
\begin{eqnarray}
\mu_{1}&=&\frac{1}{2} \left(\frac{1}{a_1^2}+\frac{1}{b_1^2}\right)+\frac{\gamma_1 {\cal N}_1}{2\pi a_1b_1}+V_1 e^{-
a_1^2}\nonumber\\&+&
\frac{\gamma_{12}{\cal N}_2}{\pi \sqrt{(a_1^2+a_2^2)(b_1^2+b_2^2)}},
\label{eq13}
\end{eqnarray}
and
\begin{eqnarray}
\mu_{2}&=&\frac{1}{2} \left(\frac{1}{a_2^2}+\frac{1}{b_2^2}\right)+\frac{V_2{\cal N}_2}{2\pi a_2 b_2} e^{-
b_2^2/{2}}\nonumber\\&+&
\frac{\gamma_{12}{\cal N}_1}{\pi \sqrt{(a_1^2+a_2^2)(b_1^2+b_2^2)}}+\frac{\gamma_2 {\cal N}_2}{2 \pi a_2 b_2}.
\label{eq14}
\end{eqnarray}
From the above equations one can calculate the  derivatives $d{\cal N}_j/d\mu_j$ and then, from the VK criterion,   
determine the stability of the  soliton. This is shown in Fig.\ref{fig2} for a specific choice of the parameters (see below).  

In the following two sub-sections we consider in more detail the cases of isotropic and anisotropic 2D binary 
solitons and compare the VA analytical predictions with numerical direct GPE time integrations.

%%%%%%%%%%%%%%%%%%%%%%%%%%%%%%%%%%%%%%
\subsection{Isotropic 2D binary solitons}
To understand the behavior of BEC profile embedded in LOL and NOL we first consider the  symmetric case $b_i=a_i$,  
$i=1,\,2$ but with $a_1\neq a_2$, giving  rise to isotropic component profiles in $x$ and $y$ directions. For simplicity 
in the following we fix $\gamma_2=0$, i.e. we assume the intra-species scattering length of the second component 
detuned to zero by a Feshbach resonance.

One can then show that the minimization of the effective energy $\langle \tilde E\rangle \equiv \langle 
E\rangle|_{b_i \rightarrow a_i}$, with respect to the $a_1$ variable, i.e. $\frac{\partial \langle \tilde E\rangle}
{\partial a_1}=0$, gives:
\begin{equation}
a_2= \left[\sqrt{\frac{-\gamma_{12} {\cal N}_2}{\pi (1+a_1^4 V_1 e^{a_1^2}) + \gamma_1 {\cal N}_1/4}}-1 
\right]^{\frac12} a_1.
\label{a2}
\end{equation}
while the minimization with respect to $a_2$, i.e. $\frac{\partial \langle \tilde E\rangle}{\partial a_2}=0$, 
allows to express $a_1$ in terms of $a_2$ as:
\begin{equation}
a_1=\left[\sqrt{\frac{-\gamma_{12} {\cal N}_1}{\pi + (1+a_2^2/2)e^{{-a_2^2}/{2}}  V_2 {\cal N}_2/4 
}}-1\right]^{\frac12} a_2.
\label{a1}
\end{equation}
%
%%%%%%%%%%%%%%%%%%%%%%%%%%%%%%%%%%%%%%%%%%%%%%%%%%%%%%%%%%%%%%%%%%%%%%%%
\begin{figure}
%\vskip -4.5cm
%\centerline{
%\hskip -.5cm
%\includegraphics[scale=0.47]{fig3.pdf}
%}
%\begin{figure}[h!]
\centerline{
\hskip -.22cm
\includegraphics[clip=true,scale=0.28]{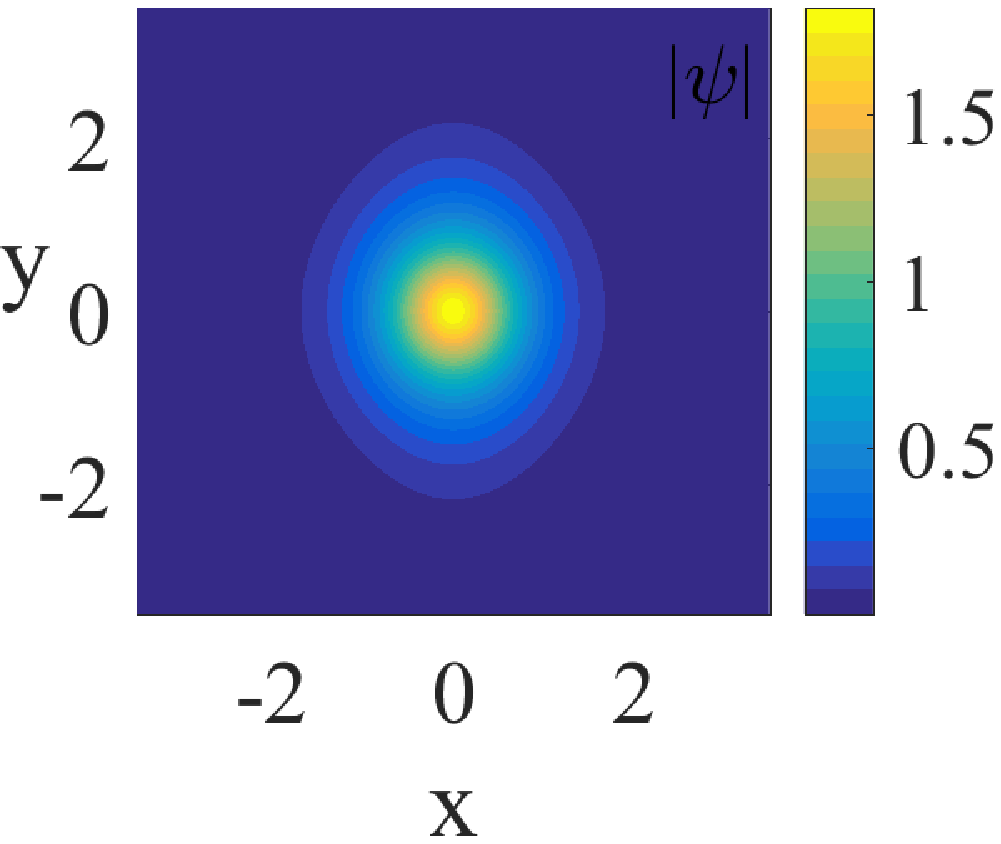}
\hskip -.2cm
\includegraphics[clip=true,scale=0.28]{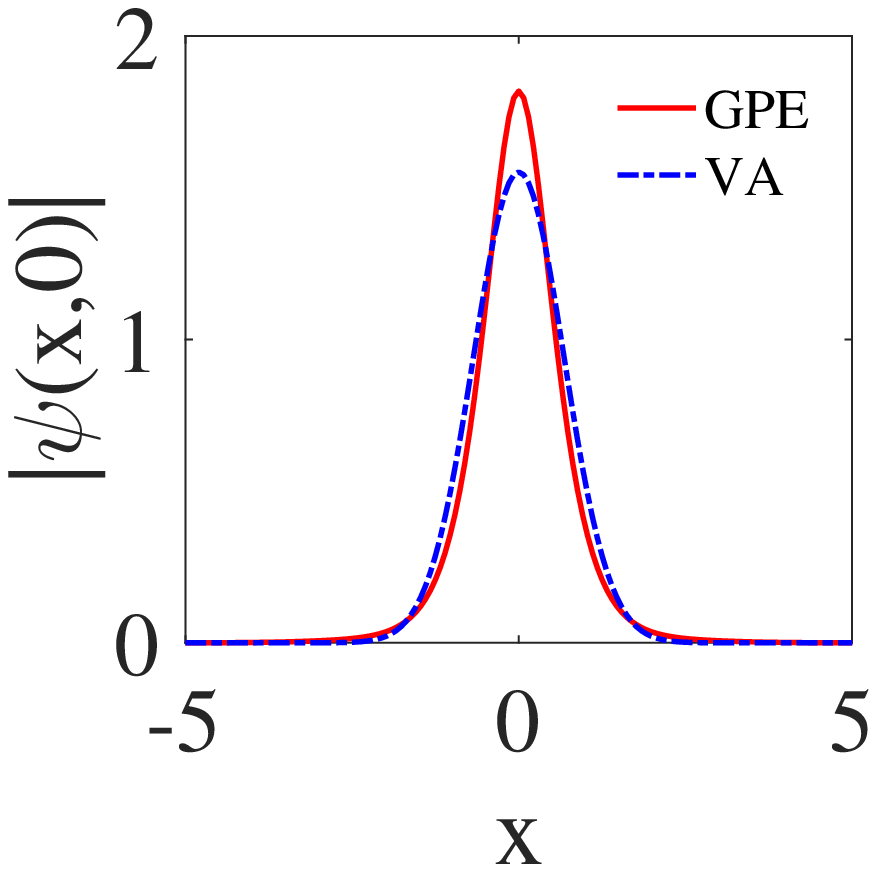}
\hskip -.24cm
\includegraphics[clip=true,scale=0.28]{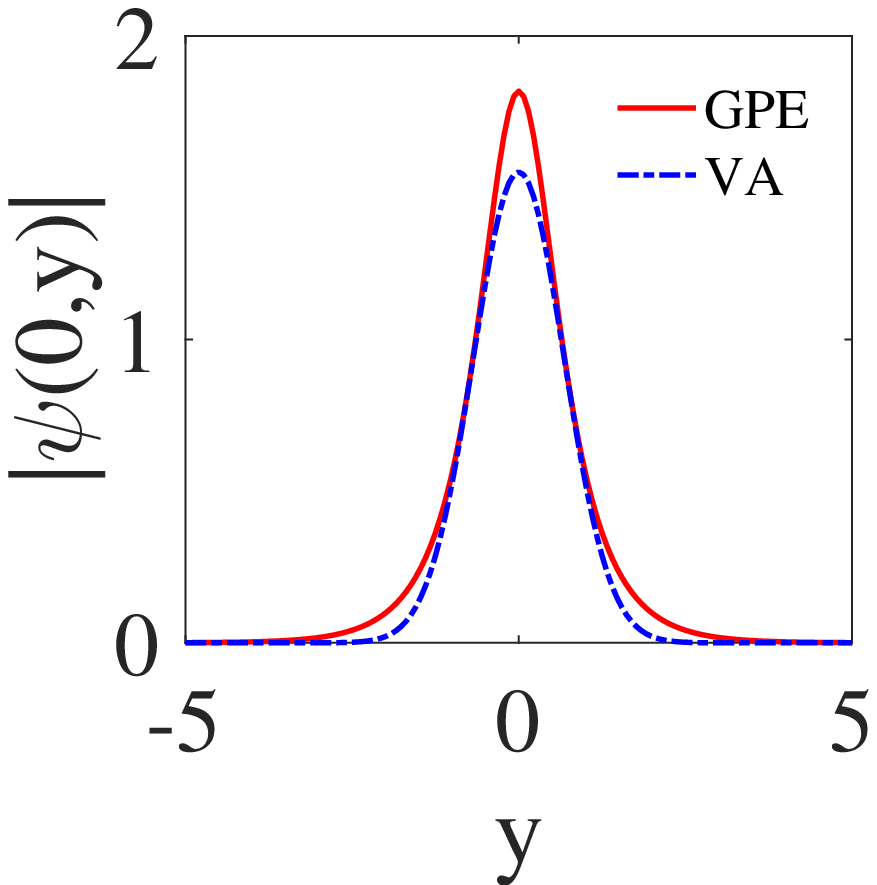}
}
\centerline{
\hskip -.22cm
\includegraphics[clip=true,scale=0.28]{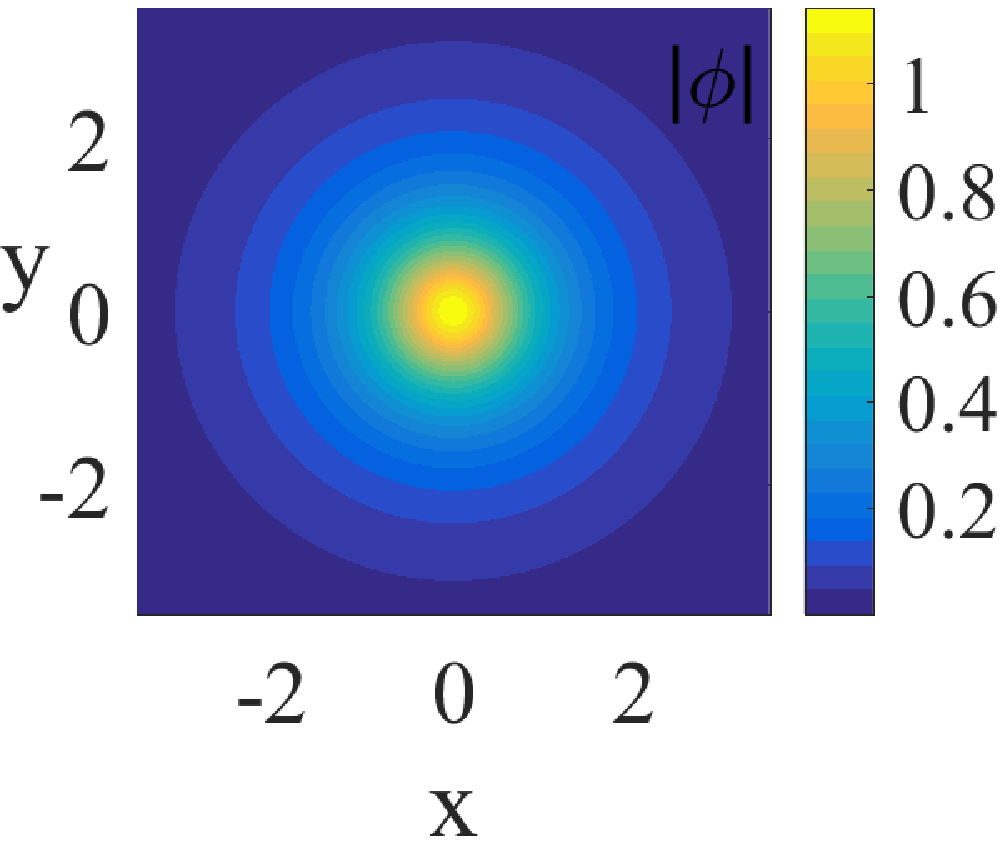}
\hskip -.2cm
\includegraphics[clip=true,scale=0.28]{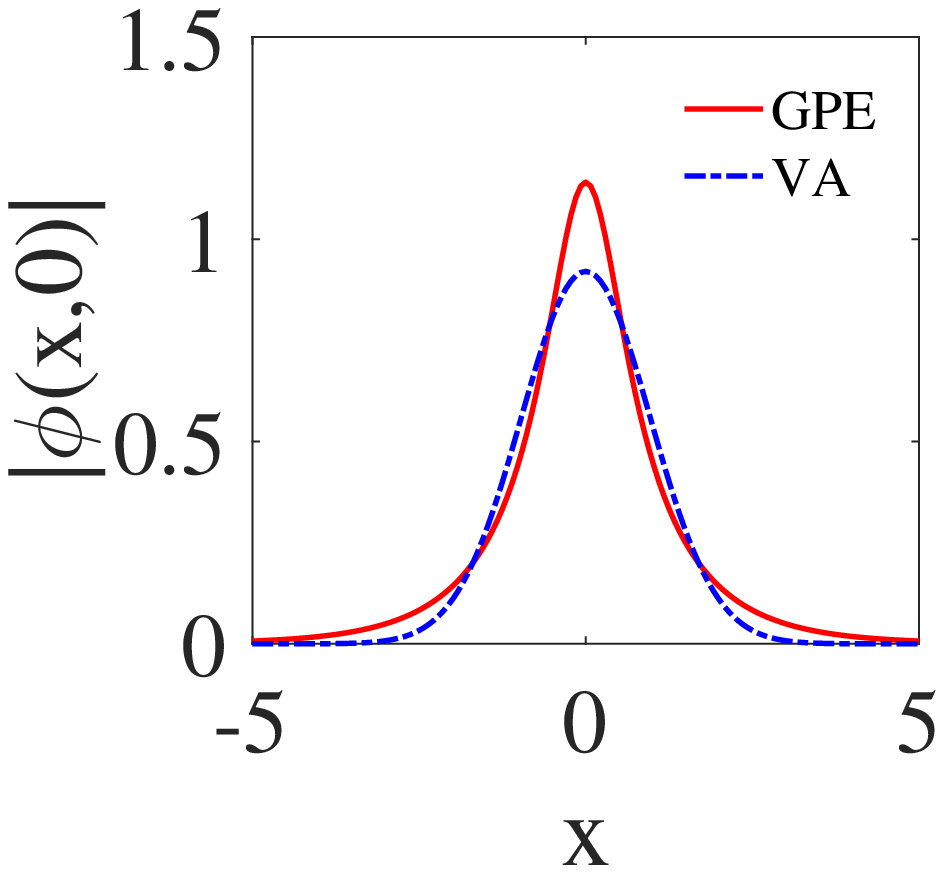}
\hskip -.22cm
\includegraphics[clip=true,scale=0.28]{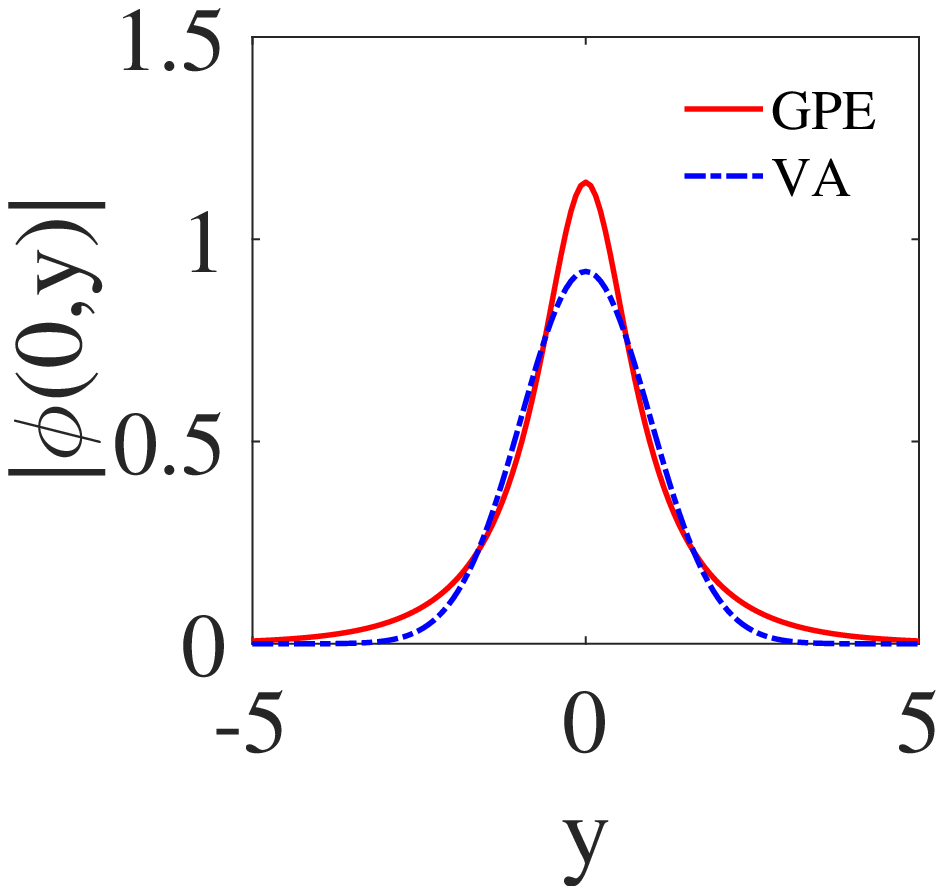}
}
%\vskip -4.5 cm
\caption{\small
2D  density plot at time  $t=100$ (left) and corresponding 1D  sections at $y=0$ (middle) and at 
$x=0$ (right) of the first (top panels) and second (bottom panels) component  of the 2D BEC solton. 
Parameters values are fixed as: $V_1=-2.0, V_2=0,\, \gamma_1=-2.0,\, \gamma_2=0,\, \gamma_{12}=-2.0,\, 
{\cal N}_1=3.5,\, {\cal N}_2=2.5$. The initial numerical profiles (red lines) overlap the ones at time $t=100$. Blue lines 
refer  to the initial profiles predicted by the VA, plotted for comparison.}
\label{fig3}
\end{figure}
%%%%%%%%%%%%%%%%%%%%%%%%%%%%%%%%%%%%%%%%%%%%%%%%%%%%%%%%%%%%%%%%%%%%%%%

Note that both equations (\ref{a2}) and (\ref{a1}) allow energy to be expressed in terms of a single variable. Due to the different 
nature of linear and nonlinear OLs, however, the two minimization cannot be performed together and one must choose one or the 
other. In fact, it can be shown that the compatibility condition of (\ref {a1}),~(\ref{a2}) is in general not satisfied for generic 
values of the parameters~\cite{note}. Since $a_2$ is the variable of the component exposed to the action of NOL, whose stability is 
more critical, it is natural to choose to minimize with respect to $a_2$, i.e. Eq.~(\ref{a1}). This is corroborated by the fact 
that by inserting the expression of $a_1$ into the energy and plotting it as a function of $a_2$ one obtains bounding   potential 
curves (see Fig.~\ref{fig1}) while the other option would lead to metastable or anti-bounding curves.

In the limit $\gamma_{12}\rightarrow 0$,  Eqs.~(\ref{eq1}),~(\ref{eq2}) become uncoupled and the problem further simplifies to two independent (single component) BECs,  one loaded in a LOL and the other in a NOL, with the corresponding  energy vs $a_2$ curves denoted as  $\langle E_{L} \rangle$ and $ \langle E_{NL} \rangle$,  respectively. In the left panel of Fig.~\ref{fig1} we show the dependence of these energies on the parameter $a_2$ both for the coupled and uncoupled cases. Notice the resemblance of these curves with the energy-potential curves  of a diatomic molecule, with $a_2$ playing the role of inter-atomic distance. From these curves it is clear that while the single component 2D BEC in the  1D LOL is energetically stable, i.e. the BEC mixture is in a bonding state (blue dashed curve), as expected from the results in Ref.~\cite{2DBEC}, no 2D soliton can be formed in a single component BEC  with NOL  since  the system is in anti-bonding state (red dash-dotted curve) when $\gamma_{12}=0$. 
\begin{figure}
\vskip -.0cm
%\centerline{
%\hskip -.5cm
%\includegraphics[scale=0.47]{fig4.pdf}
%\includegraphics[trim=2cm 0cm 1cm 0cm,clip=true,scale=0.48]{fig4.pdf}
%}
%\vskip -4.5cm
\centerline{
\hskip -.22cm
\includegraphics[clip=true,scale=0.28]{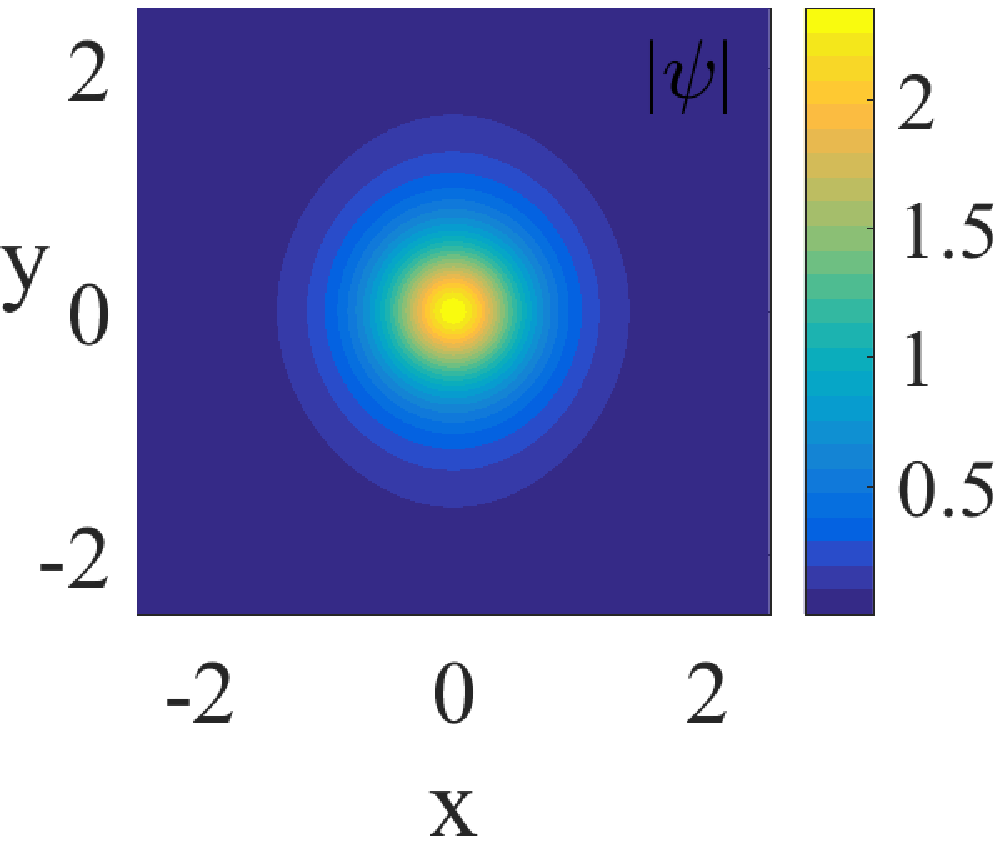}
\hskip -.2cm
\includegraphics[clip=true,scale=0.28]{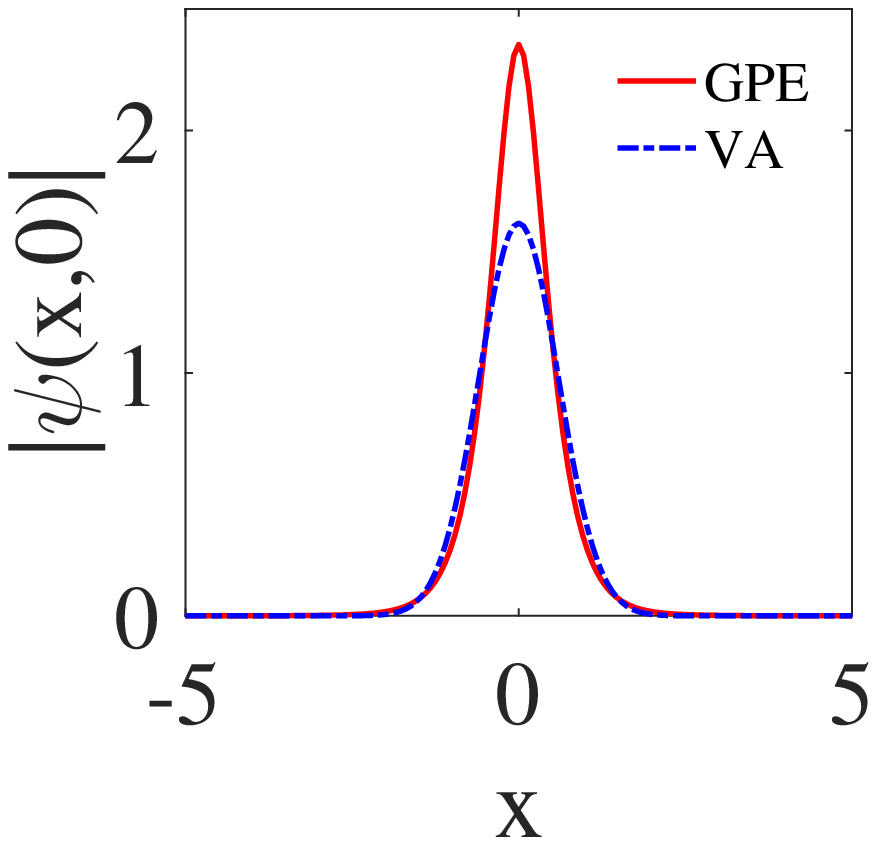}
\hskip -.24cm
\includegraphics[clip=true,scale=0.28]{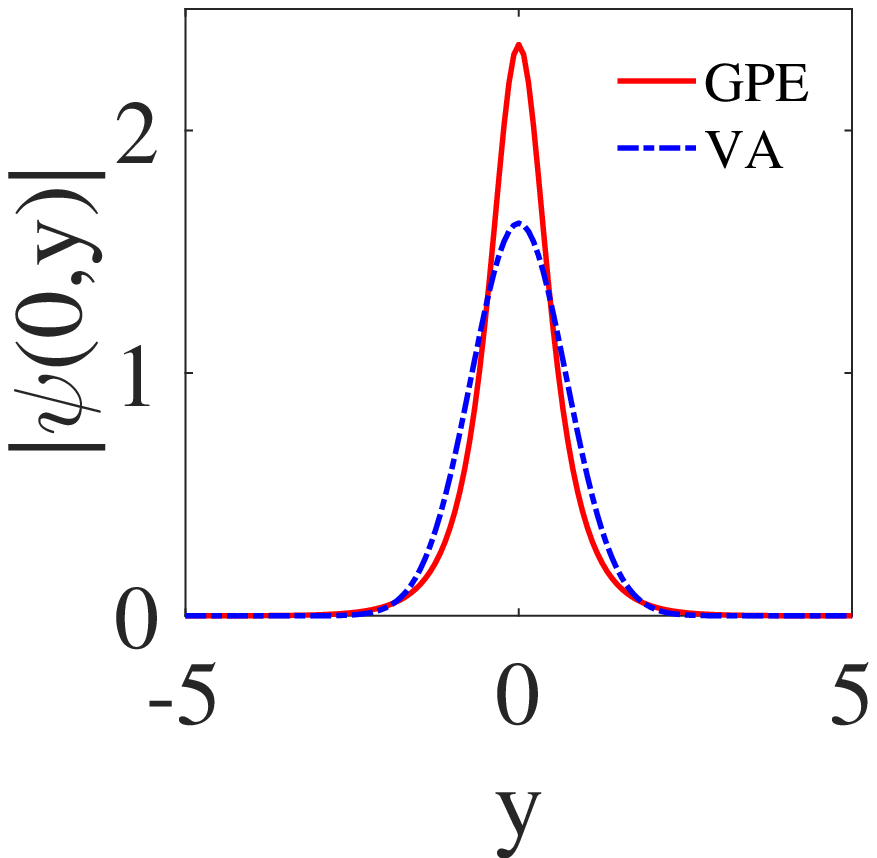}
}
\centerline{
\hskip -.22cm
\includegraphics[clip=true,scale=0.28]{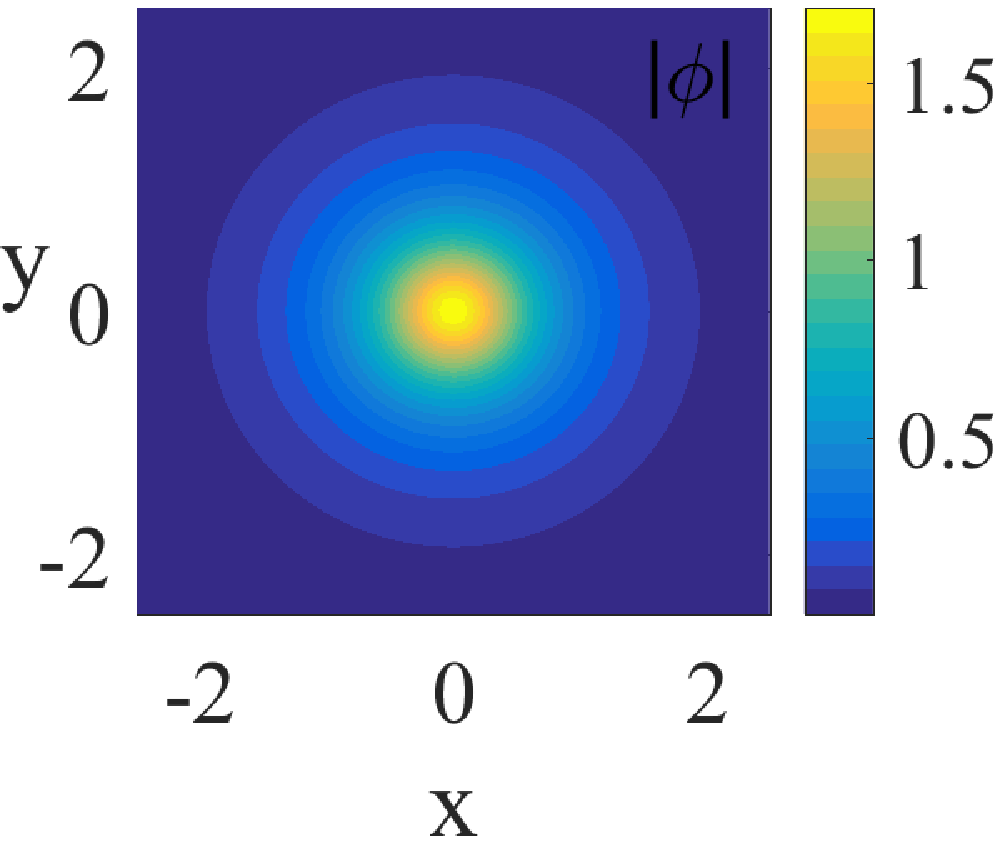}
\hskip -.2cm
\includegraphics[clip=true,scale=0.28]{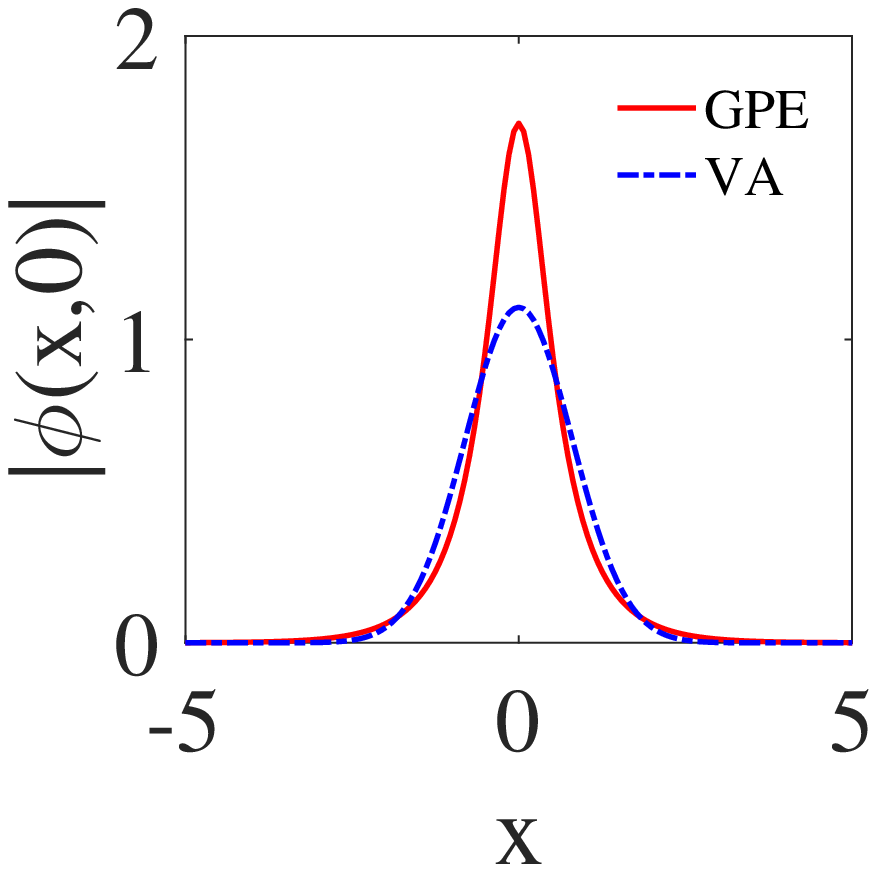}
\hskip -.22cm
\includegraphics[clip=true,scale=0.28]{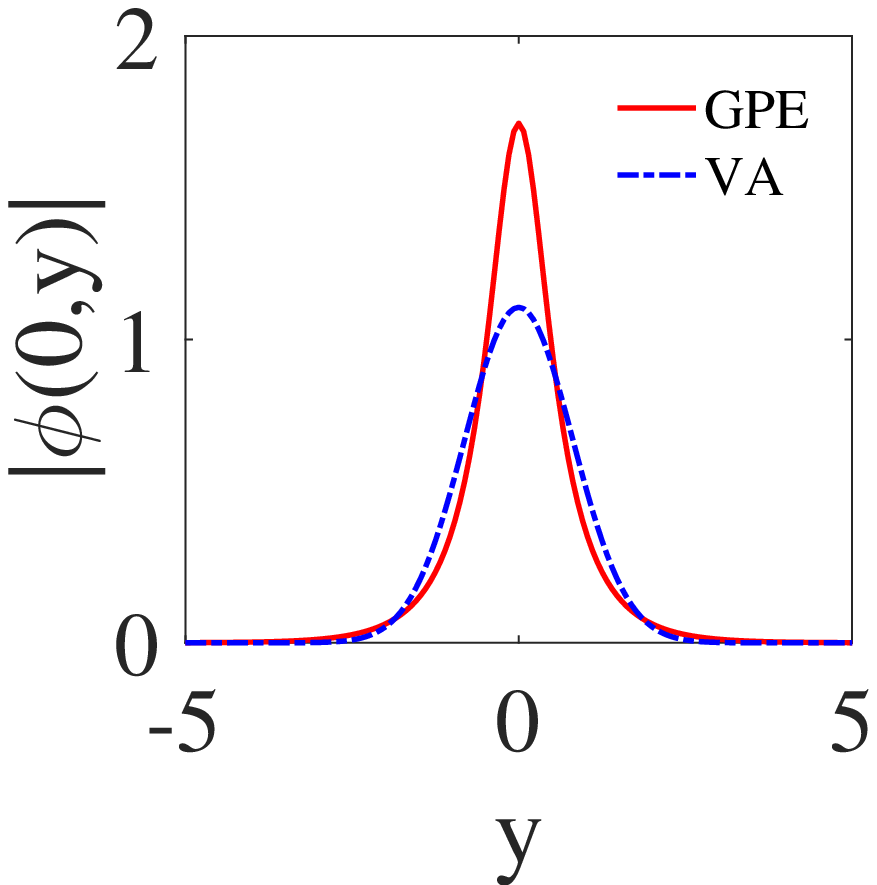}
}
\caption{\small Same as in Fig.~\ref{fig3} but for $V_2=-1$.}
\label{fig4}
\end{figure}
%%%%%%%%%%%%%%%%%%%%%%%%%%%%%%%%%%%%%%%%%%%%%%%%%%%%%%%%%%%%%%%%%%%%%
When $\gamma_{12} \neq 0$, however, the energy-potential curve develops a deeper  minimum (with respect to 
$\langle E_L \rangle$) that allows to held both components together in  a bound state (black solid curve).
This makes the components of the 2D solitons intrinsically interdependent since the presence of the inter-component 
attraction and the stability of the first component are essential for the localization of the second component and for the bound 
state formation. In the middle and right panels of  Fig.~\ref{fig1} we show the dependence of the   $\langle E \rangle$ vs $a_2$ 
curves for different values of $V_1$, keeping $V_2$ fixed, and vice-versa, respectively.

In particular, from the middle panel of Fig.~\ref{fig1} it is clear that for $V_1=0$ (absence of the LOL) there is 
no possibility to form any stable bound state, but as $V_1$ is decreased away from zero the potential develops a 
local minimum that becomes deeper and deeper as  $V_1$ is further decreased. This indicates that for fixed 
parameters and sufficiently strong LOL, symmetric binary BEC solitons can exist in the cross combined OLs 
system. The  minima of the effective energy ($\langle E \rangle_{min}$) for the $V_1=-1.5, -2, -3$ occur at $a_2 
\approx 0.69, 0.63$ and  $0.55$ respectively. Clearly, $\langle E \rangle_{min}$ is positive for $V_1=0$. It 
indicates that the system becomes unstable for weak LOL.

From the right panel of  Fig.~\ref{fig1} it is also clear that by increasing the strength, $V_2$, of the NOL,  
keeping fixed all other parameters, the local minimum of the effective potential disappears at a critical 
value  $V_{2cr}$ (for the specific choice of parameters $V_{2cr}\approx -2.53$). We find that the minimum of 
$\langle E \rangle$ occurs for $V_2=0,-1,-2$ at $a_2 \approx 0.97,\,0.75,\,0.49$ respectively. The thin black curve 
gives lower boundary of the potential curve below which the system gets collapsed ($a_1=a_2=0$ for $V_1=-2.0$ and 
$V_2=-2.52$).

We have seen that the interplay between LOL and NOL acting in two different directions plays a role in holding 
stable 2D BEC solitons. The effective energy of the system shows minima and can diverge depending on the relative 
values of the LOL and NOL. This indicates some possibilities for the observation of stable, unstable and collapse 
phenomenon. It is, therefore, constructive to check the profiles of the binary condensates from the direct 
numerical simulation of the GPEs. In particular in the following  we use the split-step Fourier (SSF) 
method~\cite{cite4} to study the spatio-temporal behaviour of BECs . The SSF is a pseudo-spectral 
method and very efficient in solving nonlinear partial differential equation~\cite{bao,hardin,shen} with very small 
time step. With  a view to calculate  the evolution of condensate in 2D  we consider a  time step $\Delta \,t=0.001$ 
and  calculate density of the BEC components. In each step, we propagate the solution by half a time step $(\Delta 
t/2)$ using nonlinear operator and then full-time step $(\Delta t)$ with linear operator and then the propagation 
is completed by the second half step $(\Delta t)$ with nonlinear operation. More explicitly,
 \begin{eqnarray}
 &&\psi(x,y,t+\Delta t)=e^{-i L_\psi \Delta t /2}\nonumber \\
 && {\cal{F}}^{-1}\left[e^{-i(k_x^2+k_y^2)\Delta t}  
{\cal{F}}\left\lbrace e^{-i L_\psi \Delta t /2} \psi(x,y,t)\right\rbrace\right]
\label{eq21}
\end{eqnarray}
and
\begin{eqnarray}
&&\phi(x,y,t+\Delta t)=e^{-i L_\phi \Delta t /2}\nonumber \\&& {\cal{F}}^{-1}\left[e^{-i(k_x^2+k_y^2)\Delta t}  
{\cal{F}}\left\lbrace e^{-i L_\phi \Delta t /2} \phi(x,y,t)\right\rbrace\right],
\label{eq22}
\end{eqnarray}
with $L_{\psi}= V_1 \cos(2 x)+\gamma_{1}|\psi|^2+\gamma_{12} |\phi|^2$ and  $L_{\phi}= 
V_{2}\cos(2y)|\phi|^2+\gamma_{2}|\phi|^2+\gamma_{12} |\psi|^2$. 
Here ${\cal{F}}$ and ${\cal{F}}^{-1}$ stand for Fourier and inverse Fourier transforms.

Clearly, Eqs.~(\ref{eq21}) and (\ref{eq22}) give time evolution of the condensate profiles for the first and second 
components respectively. In particular, dynamical stability of stationary solution is obtained from time evolution of 
initial states~\cite{golam1}. In Figs.~\ref{fig3},\ref{fig4} we plot the component densities  of the 2D binary BEC at time $t=100$, 
as obtained from numerical GPE time integration in the absence and in the presence of the NOL, respectively. We see that in both 
cases the density profiles remain stable without changing their 
norms on a long time scale. Also notice that the VA profiles (see blue lines) are in reasonably good agreement with 

the ones obtained numerically  by imaginary time evolution.

\subsection{Anisotropic 2D binary solitons}
We have seen that the consideration of isotropic condensate can provide a simplified system and allow us to 
realize system's properties in terms of potential model. However, a more general study needs to remove the 
restriction on the condensate size in different directions. Therefore, it will also be constructive to consider  
$b_i\ne a_i$, ($i=1,\,2$) and $a_1\ne a_2$.

In order to see the effect of inter-component interaction to the energy of the system, we solve Eqs.~(\ref{eq7a})-
(\ref{eq10a}) numerically and calculate minimum energy ($E_{min}$) for different values of inter-component 
interaction. The variation of $E_{min}$ with $\gamma_{12}$ is displayed in the left panel of Fig.~\ref{fig2}. It is 
seen that the  minimum  energy is negative for some non-zero values of inter-species interaction. This clearly  
indicates the existence of stable BECs.  Looking closely into the figure we see that $E_{min}$ depends sensitively 
on the interplay between optical lattices and inter-species interaction.  More specifically,  $E_{min}$ decreases 
rapidly  for larger values  of $-\gamma_{12}$. This change in $E_{min}$, however, becomes less noticeable at 
smaller values of $\gamma_{12}$. The slopes of the $\mu_j$ versus ${\cal N}_j$ displayed in the middle and right panels of 
Fig.~\ref{fig2} show that $d\mu_j/d{\cal N}_j$ is negative and thus consistent  with the VK criterion for linear stability.

With a view to study dynamical stability we  calculate the density profile by  numerically  solving Eqs.~(\ref{eq1}) and 
(\ref{eq2}) at three different times using Eqs.~(\ref{eq21})-(\ref{eq22}). We take the linearly stable stationary 
solution as initial profile and display the final profiles in Fig.~\ref{fig4}. It is clear from the time evolution 
that the coupled BECs in 2D is dynamically stable due to the interplay between non-linearity (steepening) and 
dispersive effects. 

Note that in the presence of the NOL the agreement between the VA soliton profiles and  the ones from  GPE numerical calculations 
becomes less accurate and more qualitative. This  is probably due to the fact that the Gaussian trial functions does not  allow to 
catch the full symmetry of  the cross-combined lattice. The VA stability predictions obtained with the VK criterion, however, were 
always confirmed by the numerical GPE integrations, at least for the  parameters values we have investigated.  

We finally remark that, as expected for multi-dimensional settings with attractive interactions, the above 2D binary solitons can  
undergo the collapse phenomenon. Indeed, we find that the stability depends  on norms, strength of the attractive interactions, 
strength of the OLs, and for each of these parameters there exist critical values above which the collapse occurs. The study of the 
collapse phenomenon in our setting, however, requires more detailed analytical and numerical investigations that are  out of the 
aim of this paper. 
\begin{figure}
\centerline{
\hskip -.5cm
\includegraphics[clip=true,scale=0.348]{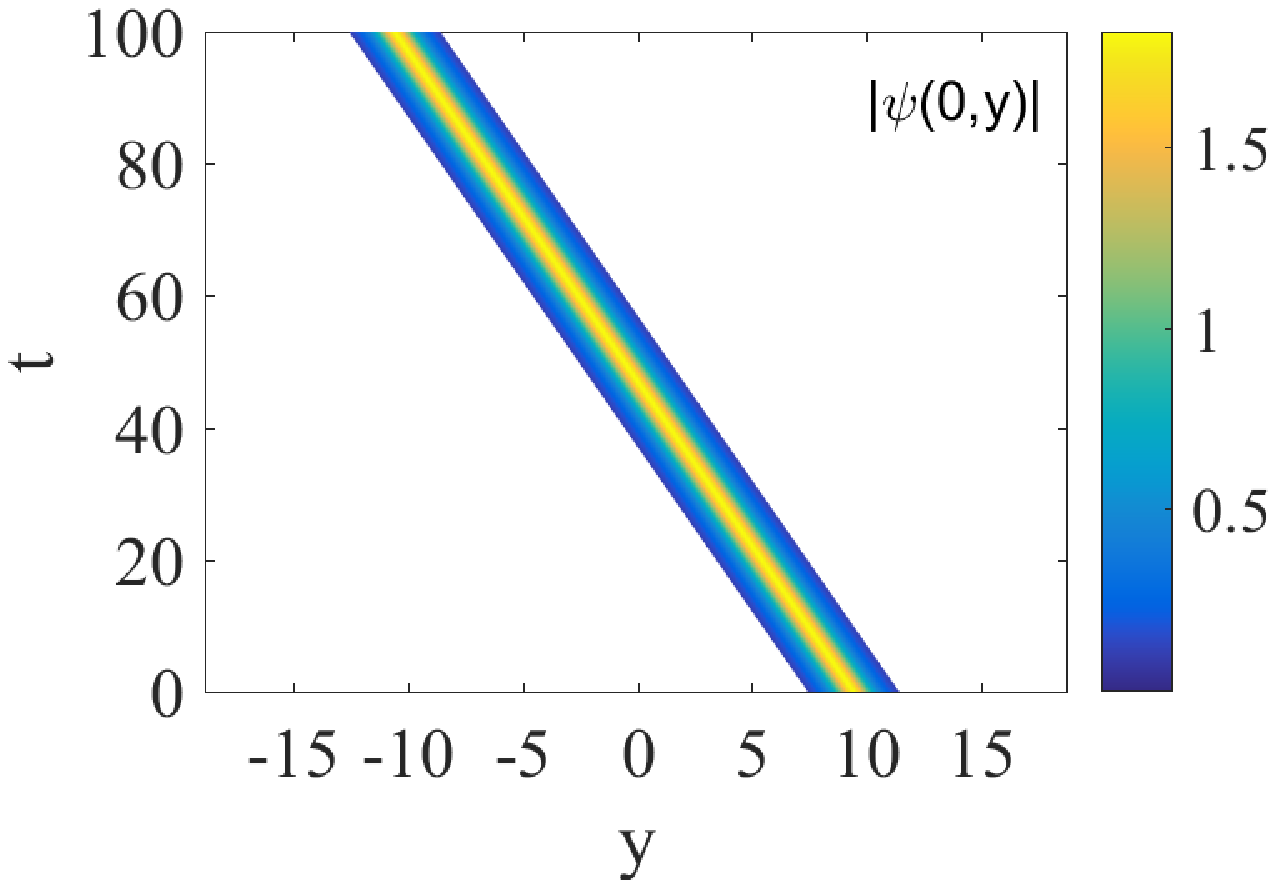}
\hskip -.25cm
\includegraphics[clip=true,scale=0.348]{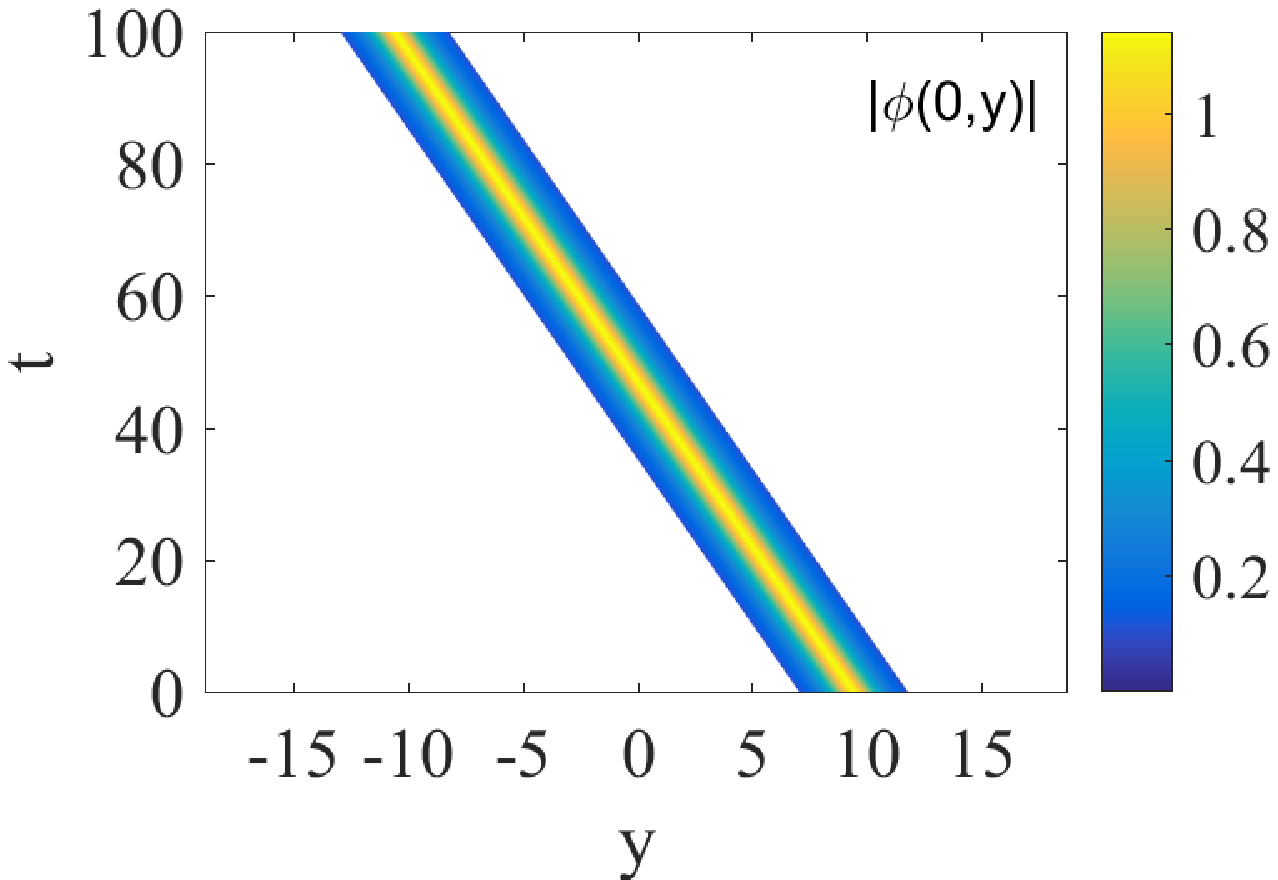}
}
\centerline{
\hskip -.5cm
\includegraphics[clip=true,scale=0.348]{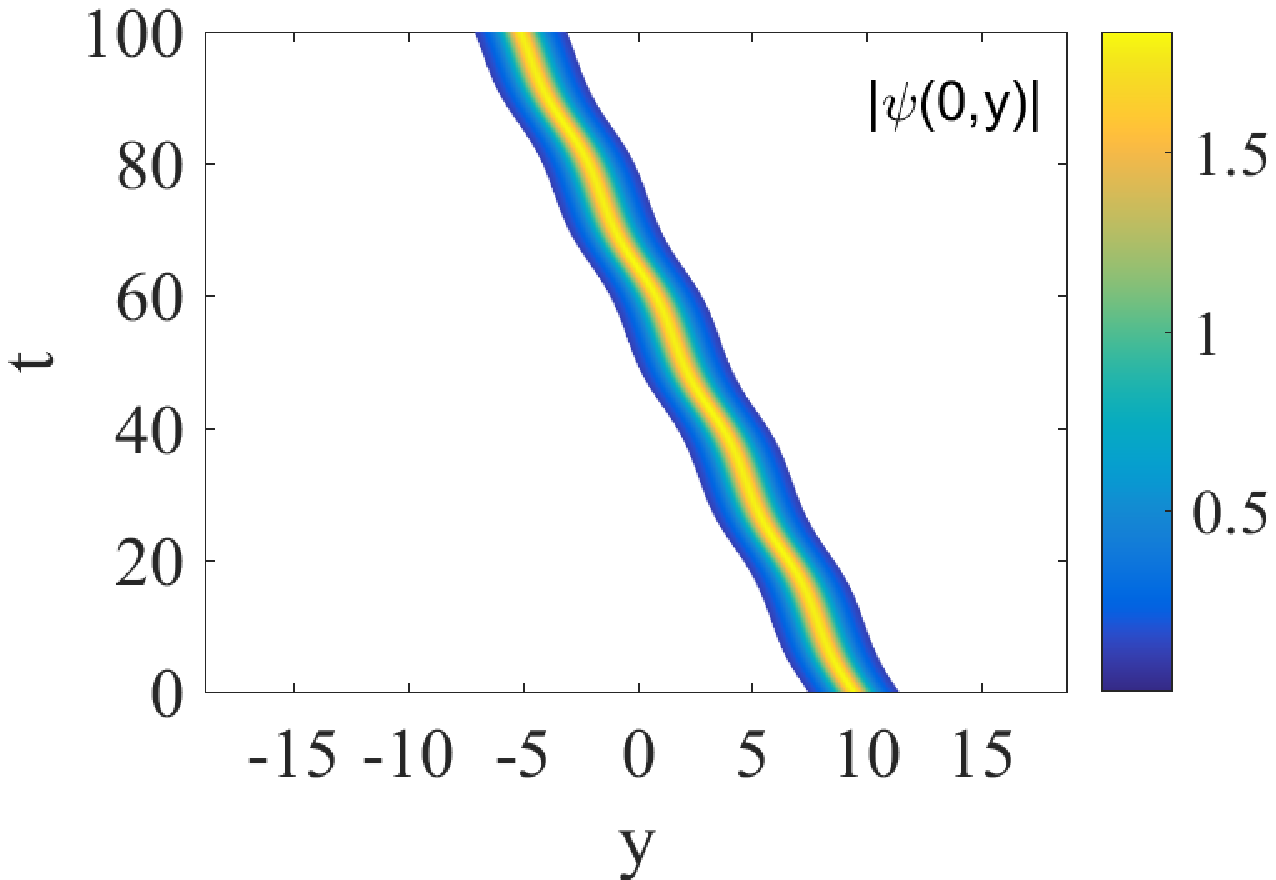}
\hskip -.25cm
\includegraphics[clip=true,scale=0.348]{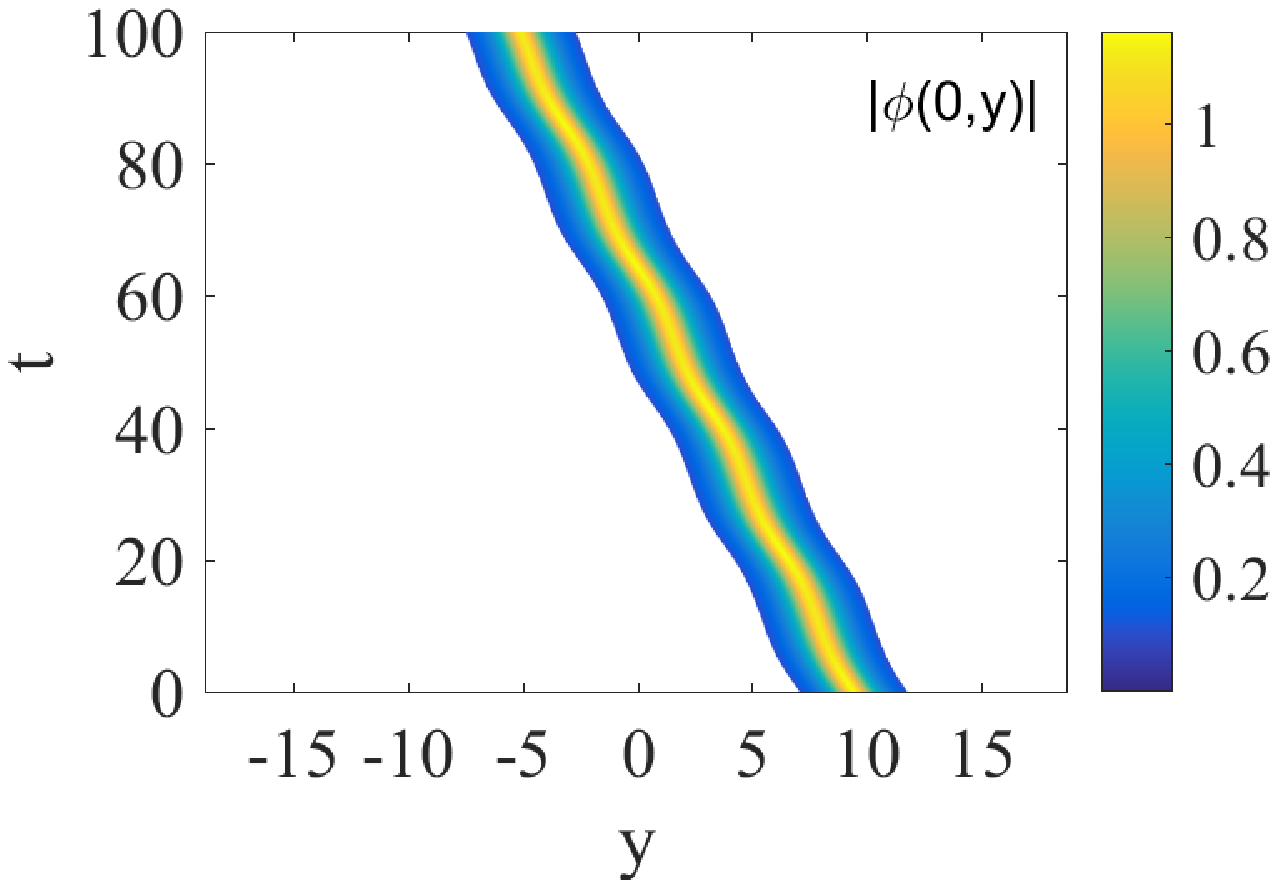}
}
\centerline{
\hskip -.5cm
\includegraphics[clip=true,scale=0.348]{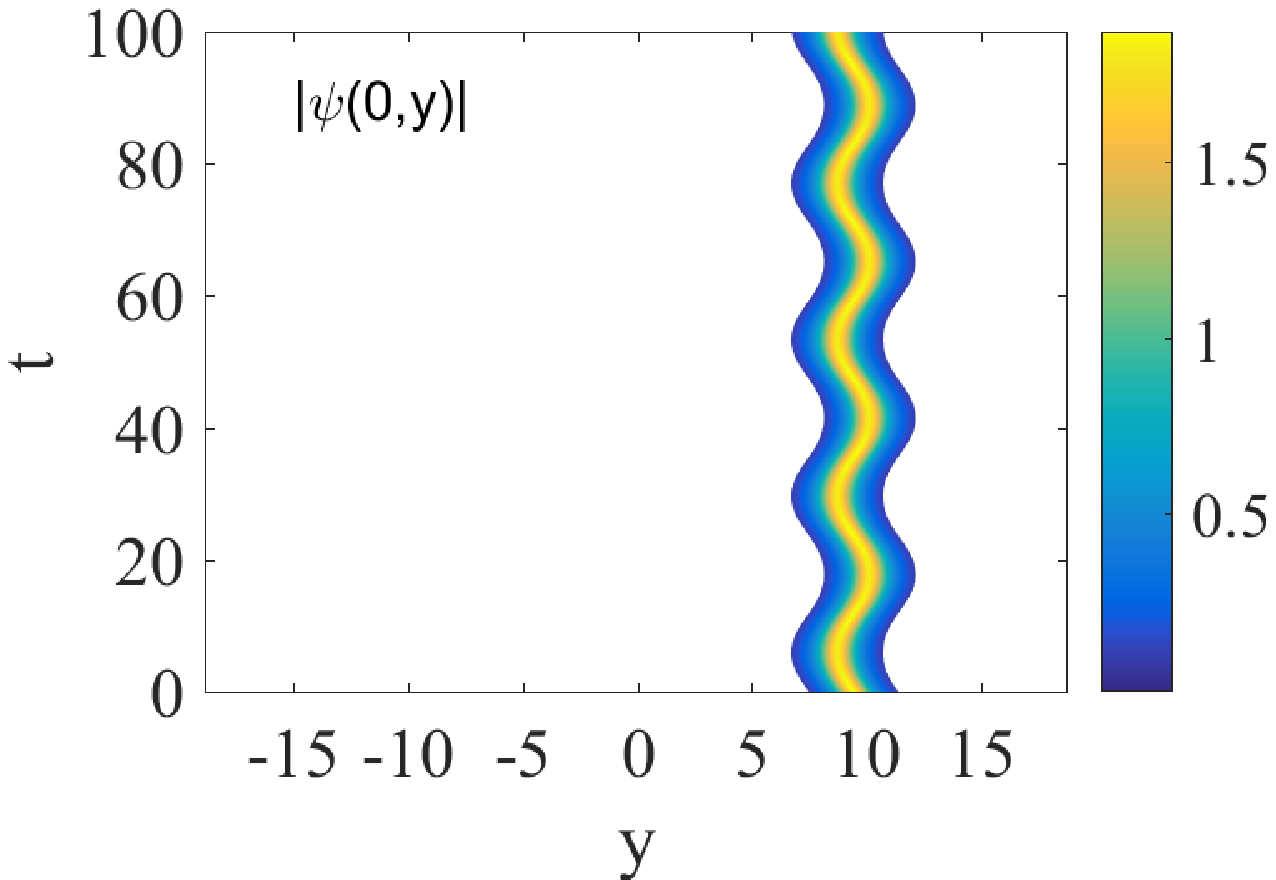}
\hskip -.25cm
\includegraphics[clip=true,scale=0.348]{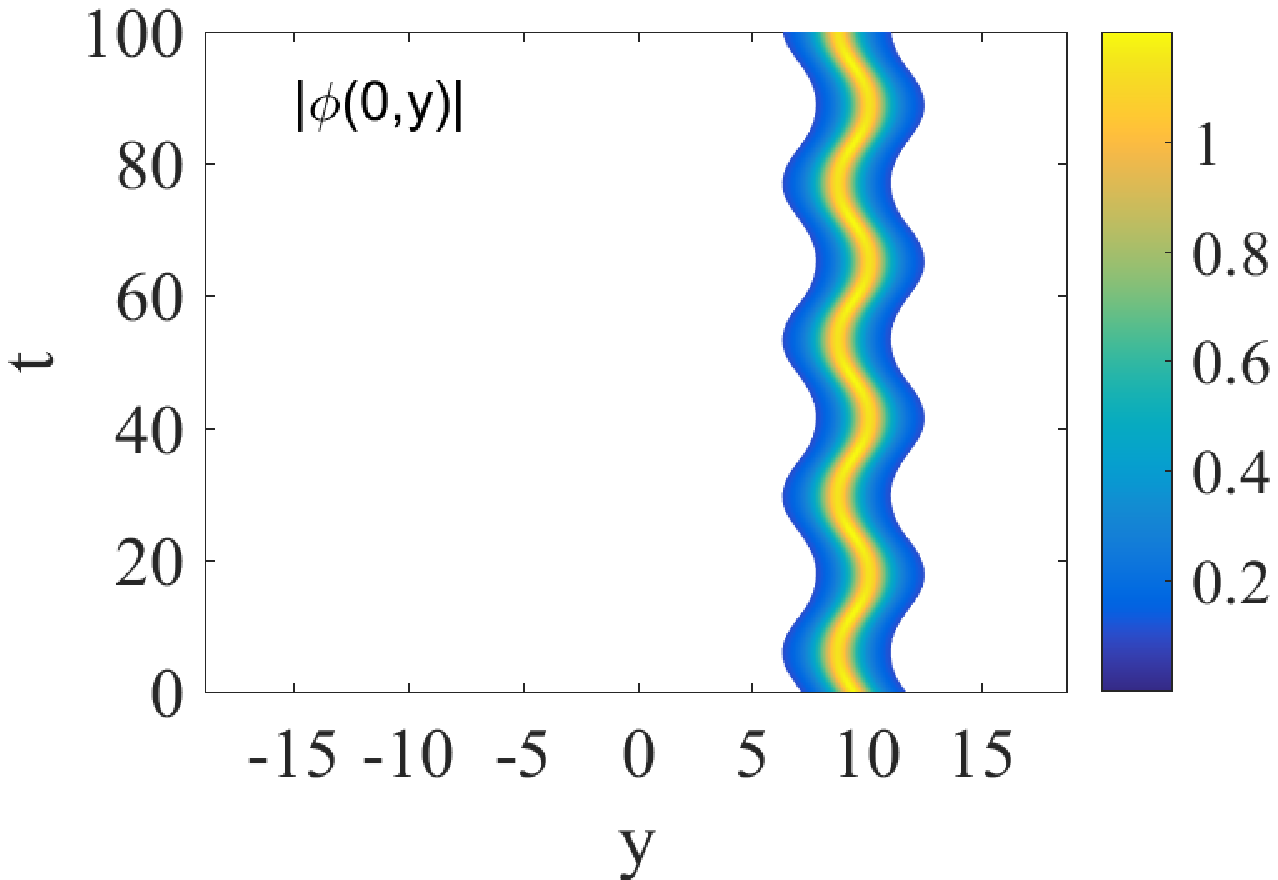}
}
\centerline{
\hskip -.5cm
\includegraphics[clip=true,scale=0.348]{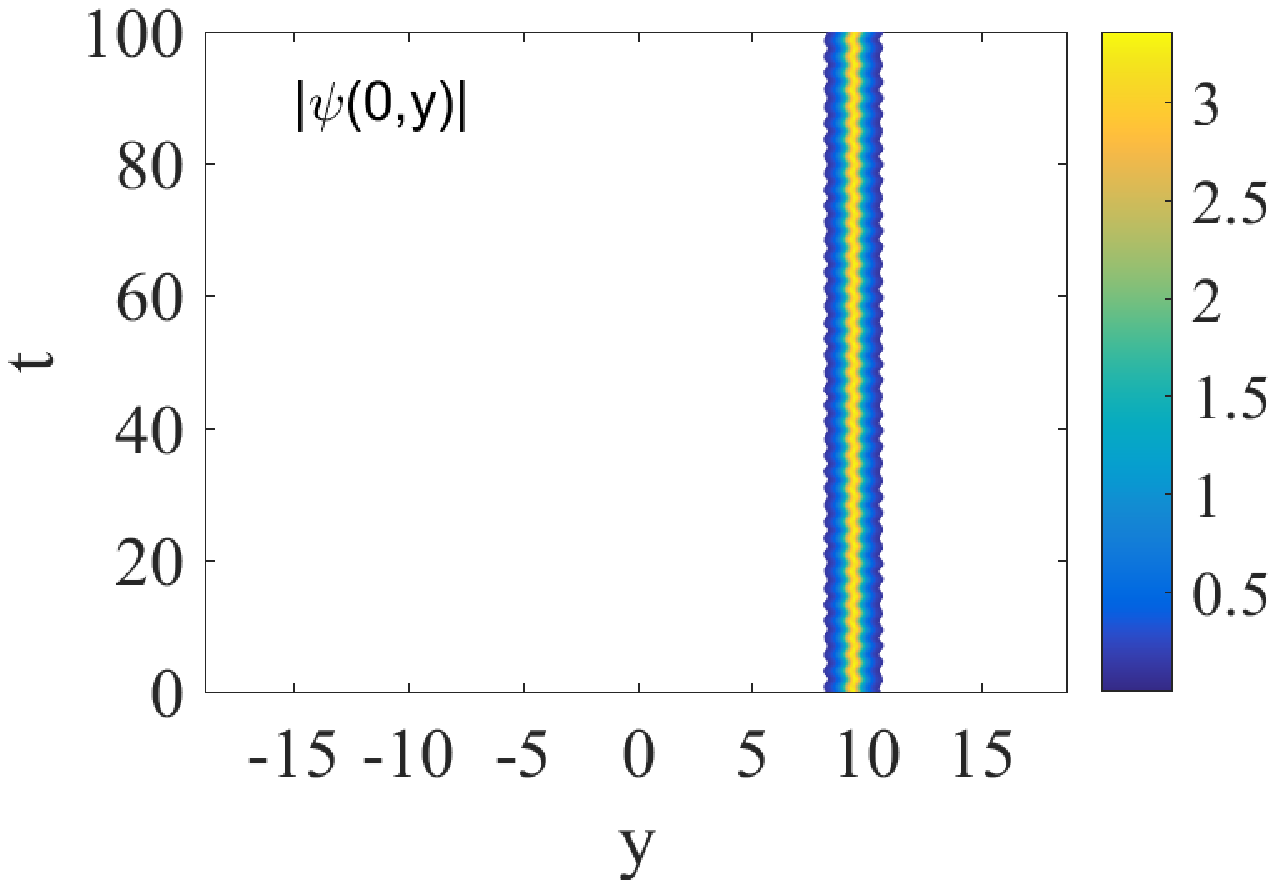}
\hskip -.25cm
\includegraphics[clip=true,scale=0.348]{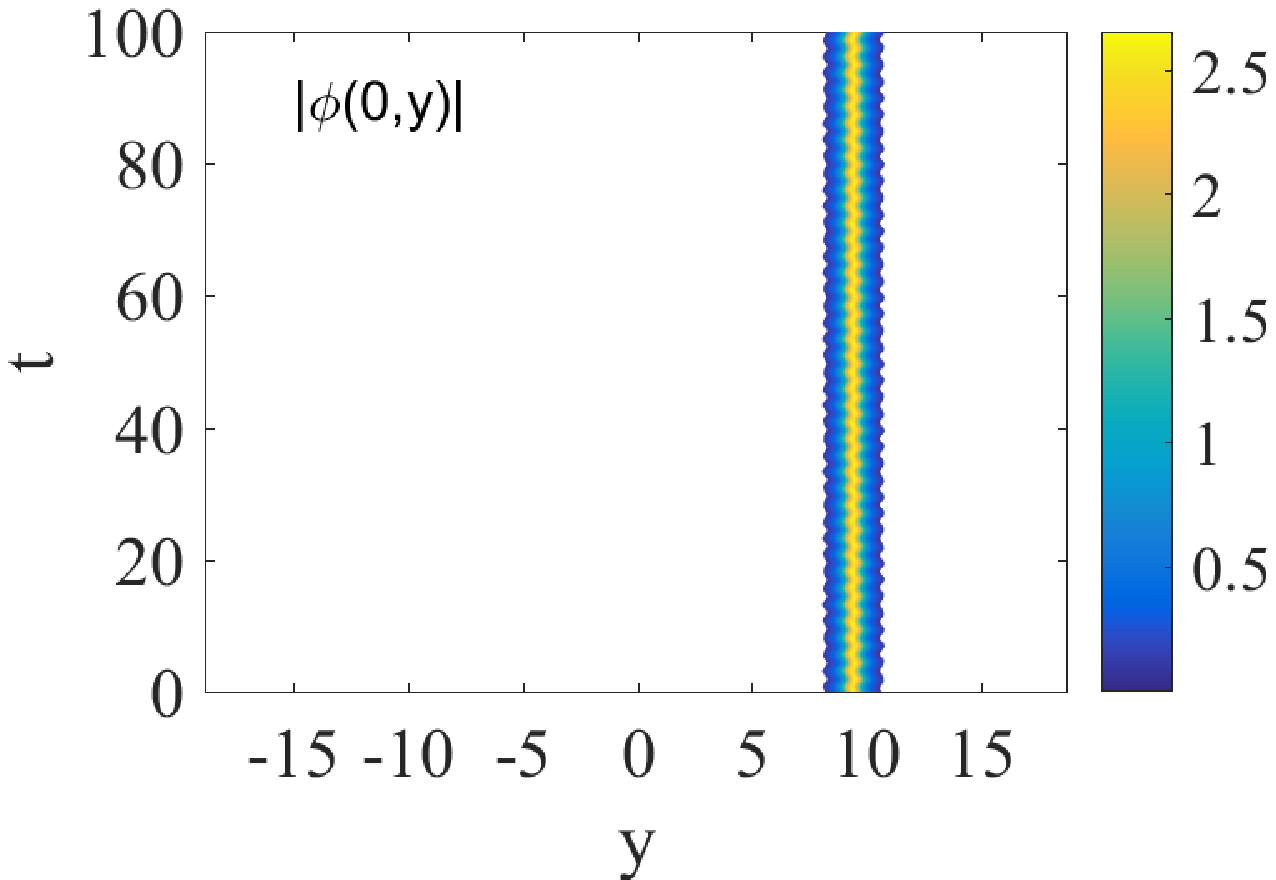}
}
\centerline{
\hskip -.5cm
\includegraphics[clip=true,scale=0.348]{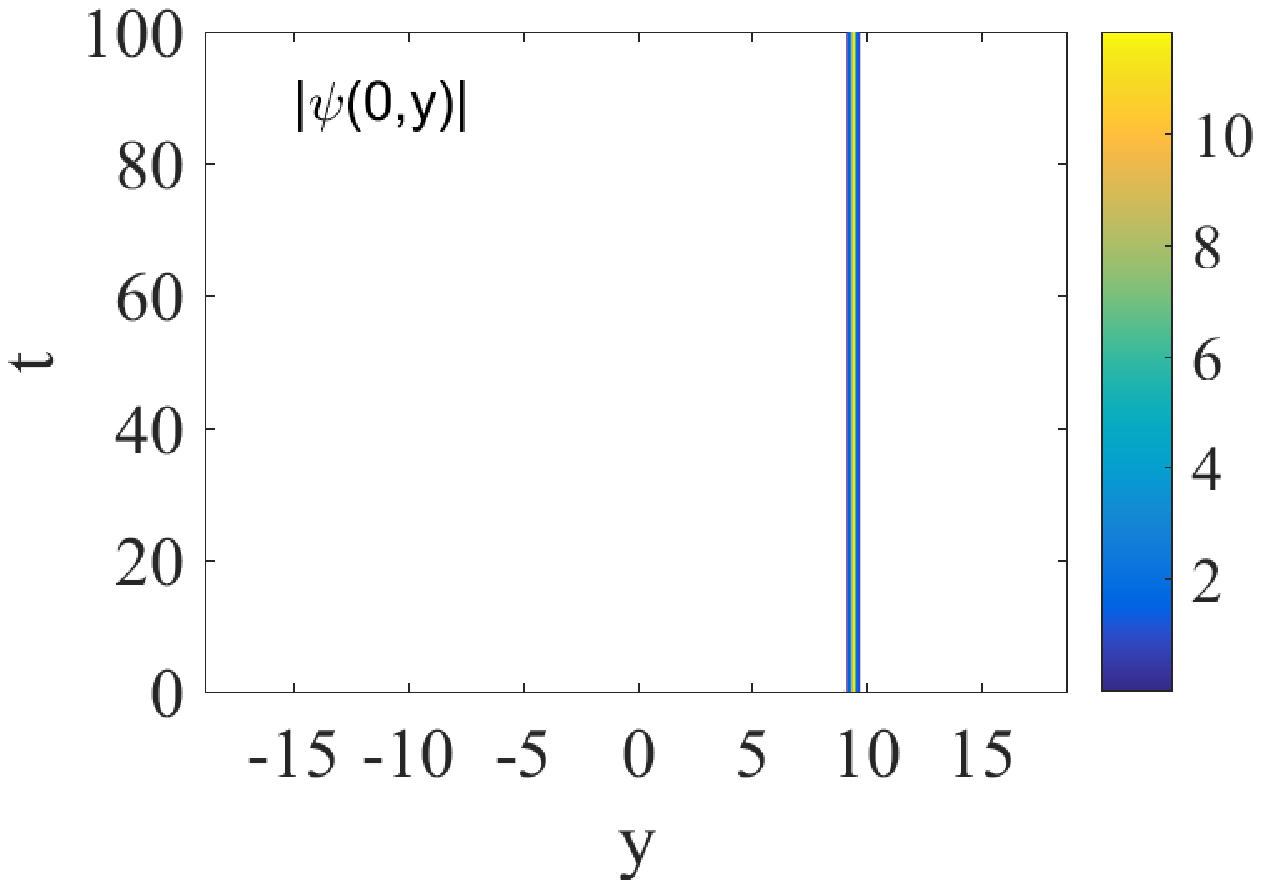}
\hskip -.25cm
\includegraphics[clip=true,scale=0.348]{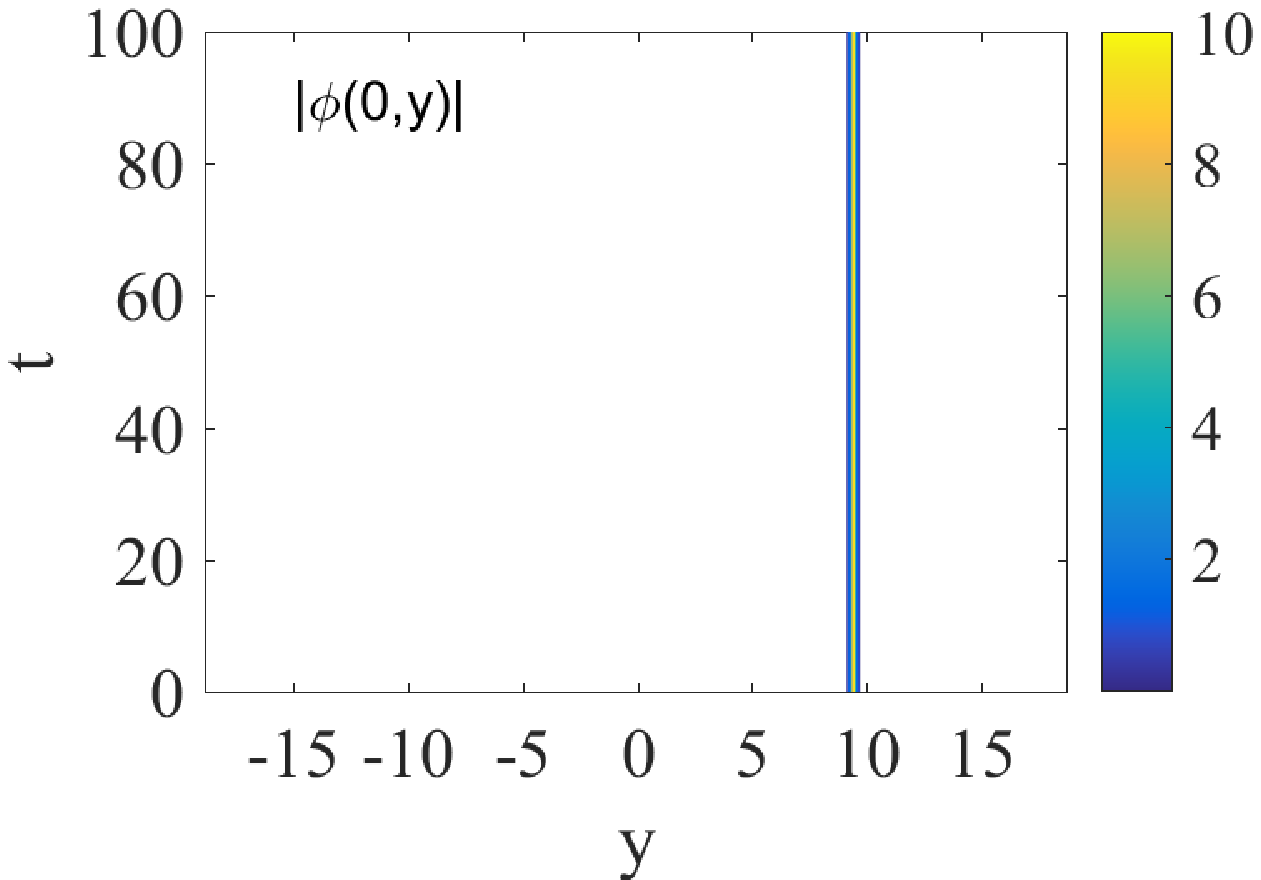}
}
\caption{\small
Density plots of the dynamics of a 2D binary soliton with phase imprinted velocity along the $y-
$direction. Panels from top to bottom refer to the different strengths of the NOL: $V_2= 0.0, -0.05, -0.15, -1.5, -2.0$, 
respectively, while left and right panels refer to the first and second component, respectively. The velocity, $v_0$, is imprinted 
by multiplying the stationary BEC components by the phase factor $\exp[-i v_0 (y- y_0)]$ with $v_0 = 0.1$ and $y_0 = 3 \pi$. Other 
parameters are fixed as: $V_1=-2.0, \gamma_{1}=-2.0$, $\gamma_{2}=0.0, 
\gamma_{12}=-2.0, N_1=3.5, N_2=2.5 $.}
\label{fig5}
\end{figure}

\begin{figure}
\centerline{
\includegraphics[clip=true,scale=0.85]{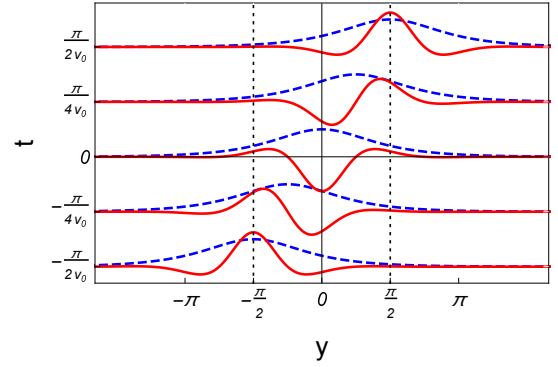}}
\caption{\small Schematic representation of the NOL effective potential in Eq. (\ref{eff-pot}) (continuous red 
lines) and of the density profiles (dashed blue lines) as functions of $y$ and at $t$.  For simplicity we assumed 
for the density y-section a solitary wave profile of the form $\frac{1}{\sqrt{2}} sech (y-v_0 t)$ and fixed 
parameters as: $v_0=0.5, V_2=-2.5$. Vertical dotted lines show the edges of the periodicity y-interval.}
\label{fig6}
\end{figure}

\section{Self-trapping action of the  NOL}
To understand the trapping action exerted by the NOL on a binary 2D BEC soliton it is convenient to consider first 
the case in which there is only the  LOL in the x direction. In this case a binary 2D matter 
wave, stabilized by the action of the 1D LOL and by the attractive inter-component interaction can freely move in 
the $y-$direction. 
This is shown in the top panels of Fig.~\ref{fig5} where an initial velocity $v_0$ in the 
$y-$direction  has been given to the stationary  components $\psi, \phi$ (obtained from imaginary time evolution) 
by means of the phase imprinting: 
\begin{equation}
\psi \rightarrow \psi e^{-i v_0 (y-y_0}, \;\; \phi \rightarrow \phi \psi e^{-i v_0 (y-y_0}.
\nonumber 
\end{equation}
This state is then used as initial condition for the real time integration of the coupled GPE system. Results are 
reported in Fig.~\ref{fig5} for a fixed imprinted velocity and  different values of the strength of the NOL. We see that in absence 
of the NOL the 2D binary matter wave moves  like a soliton retaining its shape and initial velocity,  while in the presence of the 
NOL and for a fixed imprinted velocity, the soliton can either move  or become  dynamically self-trapped (i.e oscillates around 
some positions),  depending on the strength of the NOL being  below or above a certain threshold, 
respectively. Notice that the amplitude of the period of the oscillations in the self-trapped regime decreases by increasing the 
strength of the NOL, with the binary soliton becoming fully at rest for sufficiently high values of $V_2$ (in Fig.~\ref{fig5}
this occurs for $V_2=-2.0$, as one can see from the bottom panel).

The origin of the self-trapping can be qualitatively understood by taking into account the effective potential, 
${\cal{V}}_{NOL}$, induced by the NOL on the moving second component. For simplicity we assume  for this a solitary 
wave localized in a minimum of the LOL (i.e. stationary with respect to the LOL)  moving with constant velocity $v_0$ 
in the y-direction. In this case the effective NOL potential has the form: 
\begin{equation}
{\cal{V}}_{NOL}= V_2 \cos(2 y) |\phi(x, \zeta(t)|^2,
\label{eff-pot}
\end{equation}
with $\zeta(t)=y-v_0 t$ denoting the traveling  wave coordinate.  
Notice that the potential in Eq.~(\ref{eff-pot}) is  periodic in $y$ (with period equal to $\pi$) and that while 
the density profile depends on time, the sinusoidal factor is independent of $t$, being related to the stationary 
wave that spatially modulates the intra-species interaction of the second component via an optically induced 
Feshbach resonance.

As the BEC density  moves,  the shape of the potential changes in the periodicity interval $y \in [-\pi/2, 
\pi/2]$ as schematically shown in Fig.~\ref{fig6}. We see that at $t=0$ ($y=0$) the soliton density is 
located at the  minimum of ${\cal{V}}_{NOL}$ (which has the form of a potential well) and  at 
$t=\pm \pi/2v_0$ (i.e. $y=\pm \pi/2$) it is located at the maximum of ${\cal{V}}_{NOL}$ (which has the form of a 
potential barrier). In order to move then  the soliton  must have at least the energy necessary to overcome the 
potential barriers faced at the edges of the periodicity interval of ${\cal{V}}_{NOL}$. 
%
%%%%%%%%%%%%%%%%%%%%%%%%%%%%%%%%%%%%%%%%%%%%%%%%%%%%%%%%%%%%%%%%%%%%%%%%%
\begin{figure}
\centerline{
\includegraphics[clip=true,scale=0.65]{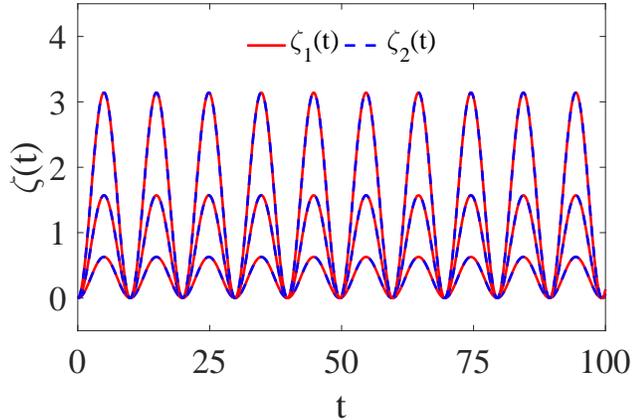}}
\caption{\small The dynamics of the COM coordinates of a binary soliton in a parabolic trap $\beta (y-y_0)^2$, in 
the absence of the NOL $(V_2 = 0)$ for different initial trap displacements: 
$y_0=0.1 \pi, 0.25  \pi, 0.5 \pi$, corresponding to curves with amplitudes ordered from smaller to larger, 
respectively. Other parameters are fixed as: $\beta=0.1, \gamma_1=-2.0, \gamma_2=0, \gamma_{12}=-2.0, V_1=-2.0, 
N_1=3.5, N_2=2.5$. Note that the COM curves of the two components are in perfectly overlapping.}
\label{fig7}
\end{figure}
%%%%%%%%%%%%%%%%%%%%%%%%%%%%%%%%%%%%%%%%%%%%%%%%%%%%%%%%%%%%%%%%%%%%%%%%%%%%
%
On the contrary, the localized matter wave remains trapped inside the NOL effective potential and oscillates around its minimum. 
This bears resemblance to the Peirls-Nabarro barrier  that discrete solitons must overcome in order to 
move~\cite{kivshar-campbell}. In our case, however, the barrier depends on the dynamics and is self created 
by the  wave through its density, for this we refer to it as {\it  dynamical self-trapping} barrier.   

As is well known, the effect of the usual Peirls-Nabarro barrier on discrete solitons is the slowing down of the 
their motion and eventually their  stopping (pinning) at some lattice site. These behaviors are  similar 
to what we observe in our numerical experiments (see below). In the presence of attractive  inter-component interaction, necessary 
for  the binary 2D soliton to exist, the stopping  of the second component implies the stopping of the first component as well,  
and therefore  the dynamical self-trapping of the 2D binary soliton. This qualitatively explains  the physical  mechanism by which 
the DSTP arises in the presence of the NOL~\cite{note2}.

\section{DSTP in external potentials}
Dynamical  behaviours similar to the ones of the previous section are expected  for  binary BEC solitons  put in 
action by external potentials. In the absence of the NOL (i.e. with only the LOL  in the $x-$direction),  we 
find (see below) that the addition of a parabolic trap in the y-direction makes the soliton oscillating around 
the minimum of the potential, while the addition of a ramp potential produces a uniform acceleration along 
the $y$-direction,  just as one would expect for ordinary solitons. 
%
%%%%%%%%%%%%%%%%%%%%%%%%%%%%%%%%%%%%%%%%%%%%%%%%%%%%%%%%%%%%%%%%%%%%%%%
\begin{figure}
\centerline{
\includegraphics[clip=true,scale=0.65]{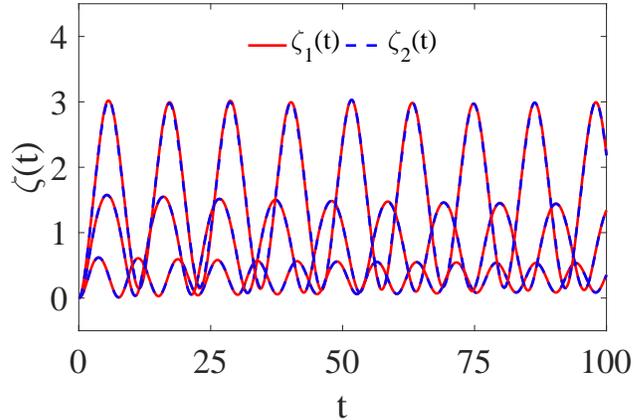}}
\caption{\small Same as in Fig.\ref{fig7} but for $V_2 = -0.5$ and initial displacements $y_0=0.2\pi,\,  0.35\pi, 
\, 0.5\pi $.}
\label{fig8}
\end{figure}
%%%%%%%%%%%%%%%%%%%%%%%%%%%%%%%%%%%%%%%%%%%%%%%%%%%%%%%%%%%%%%%%%%%%%%%%%
%

The presence of the NOL in the $y$-direction, however, introduces  the possibility of self-trapping phenomena intuitively expected 
when the soliton energy is not enough for the  overcoming of the effective NOL barrier discussed before. 
This automatically  implies the existence of thresholds in the parameter space as  demonstrated in the following sub-
sections for the specific cases of  parabolic traps and linear ramp potentials. 

To show this we recourse to numerical integration of the coupled GPE system taking binary solitons 
obtained from imaginary time evolution as initial conditions. The potentials are applied in all cases along  the $y-$direction  and 
the center of mass of the initial conditions are located at  the $(x=0, y=0)$ position in the cross-combined potential. 
With this setup the motion of the binary soliton along the x-axis is strongly confined by the LOL to the $x = 0$ 
channel and the dynamics mainly occurs in the y-direction (this is particularly true for deep LOL and strong inter-
component interactions). The dynamics is then investigated in terms of the center of mass (COM) of the two BEC 
components defined by:
\begin{equation}
\zeta_j(t)=\frac{\int\!\!\!\int^{\infty}_{-\infty}y\,|\psi_j|^2dxdy}{\int\!\!\!\int^{\infty}_{-\infty}|\psi_j|^2dxdy}, \quad j=1,2.
\end{equation}
\begin{figure*}
\centerline{
\hskip -.25 cm
\includegraphics[scale=0.38]{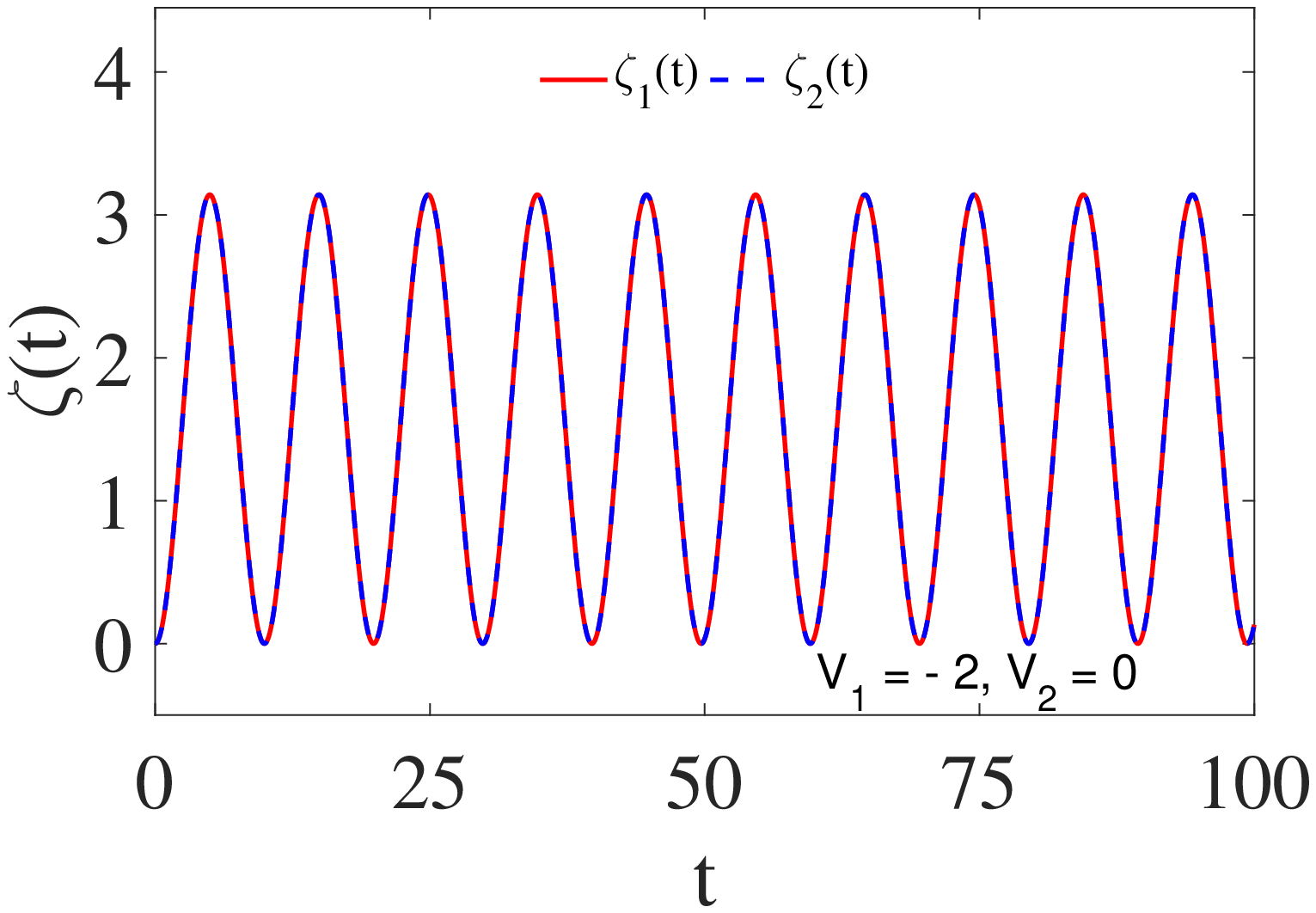}
\hskip -.45cm
\includegraphics[scale=0.38]{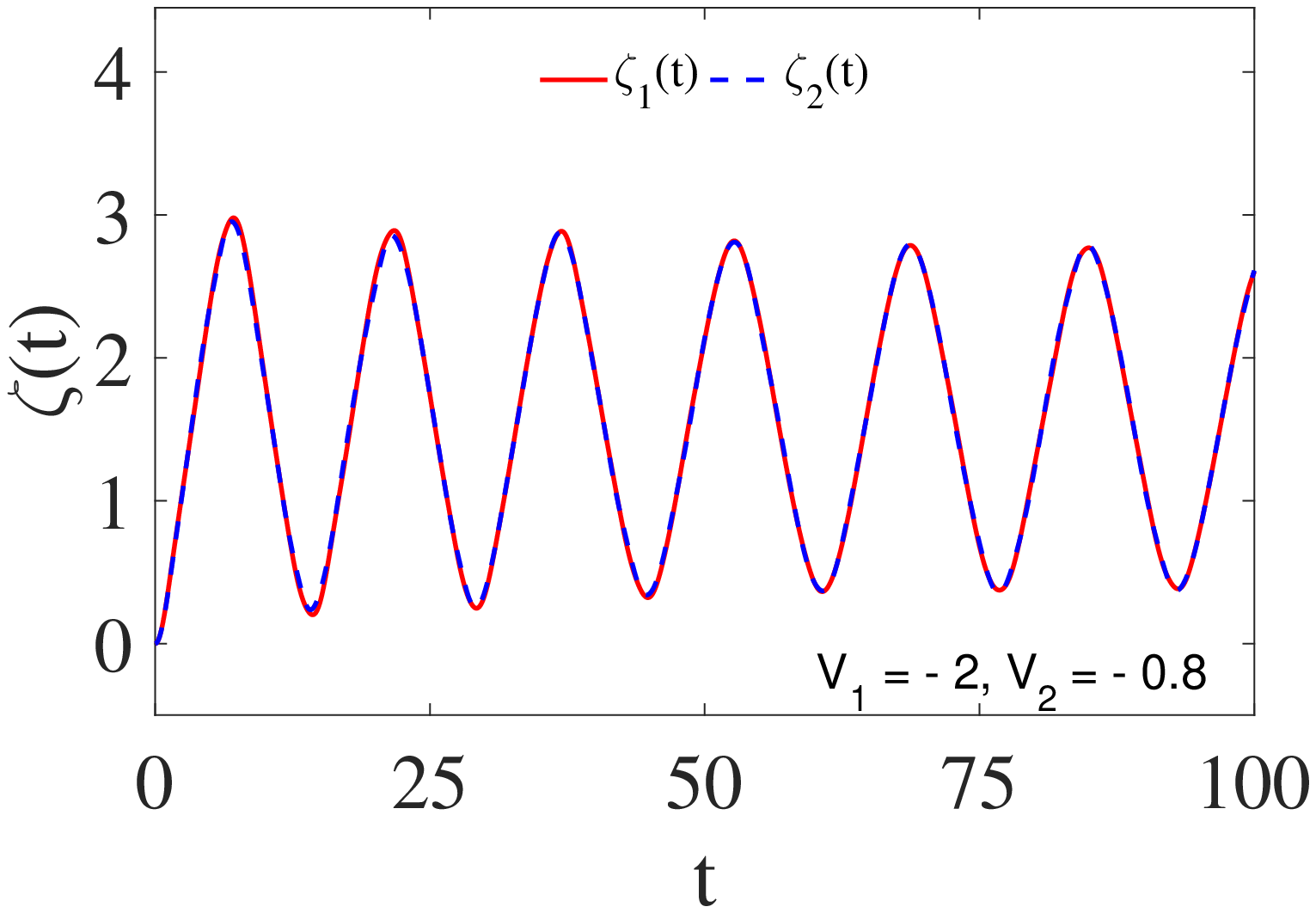}
\hskip -.45cm
\includegraphics[scale=0.38]{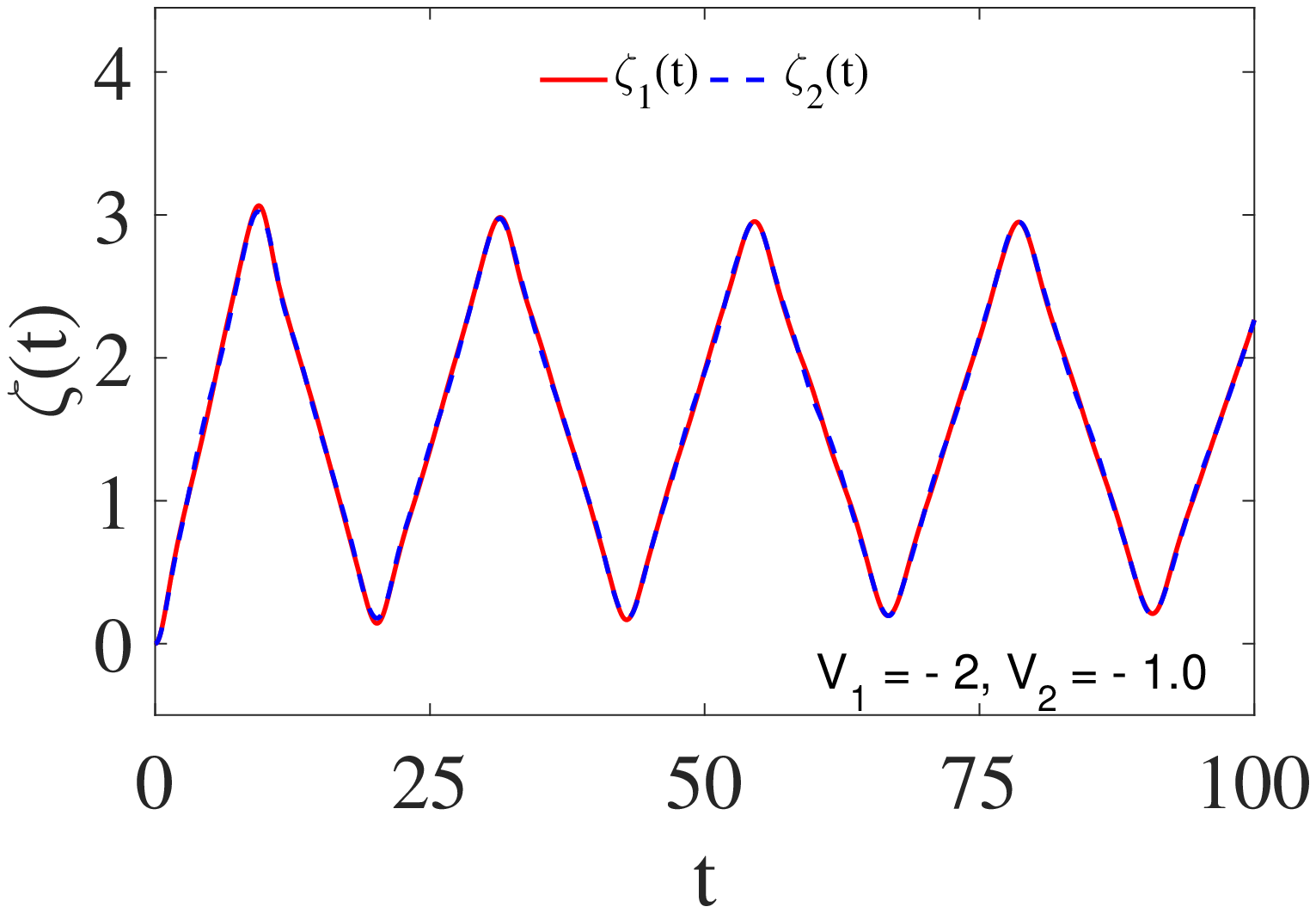}}
\centerline{
\includegraphics[scale=0.38]{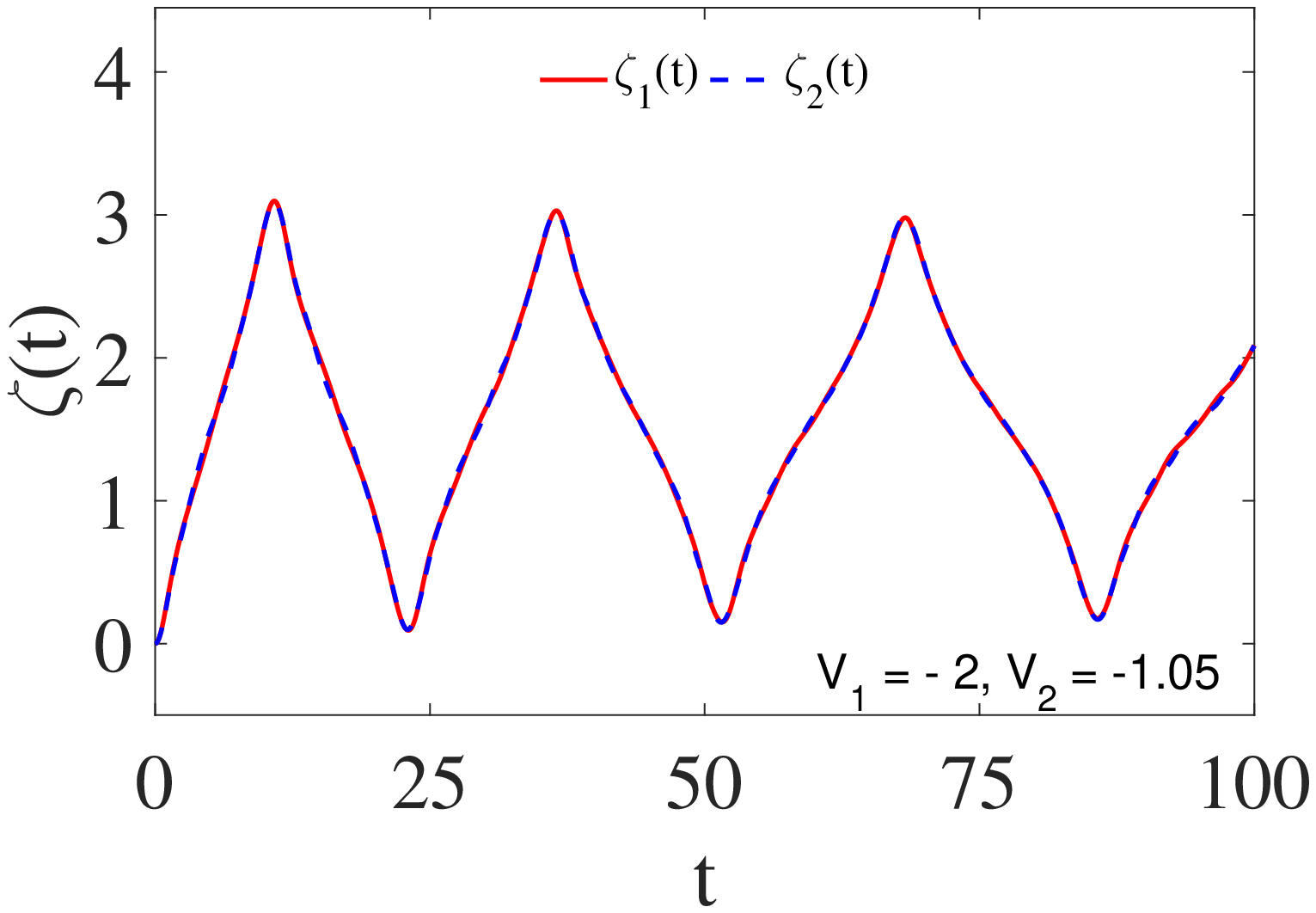}
\hskip -.45cm
\includegraphics[scale=0.38]{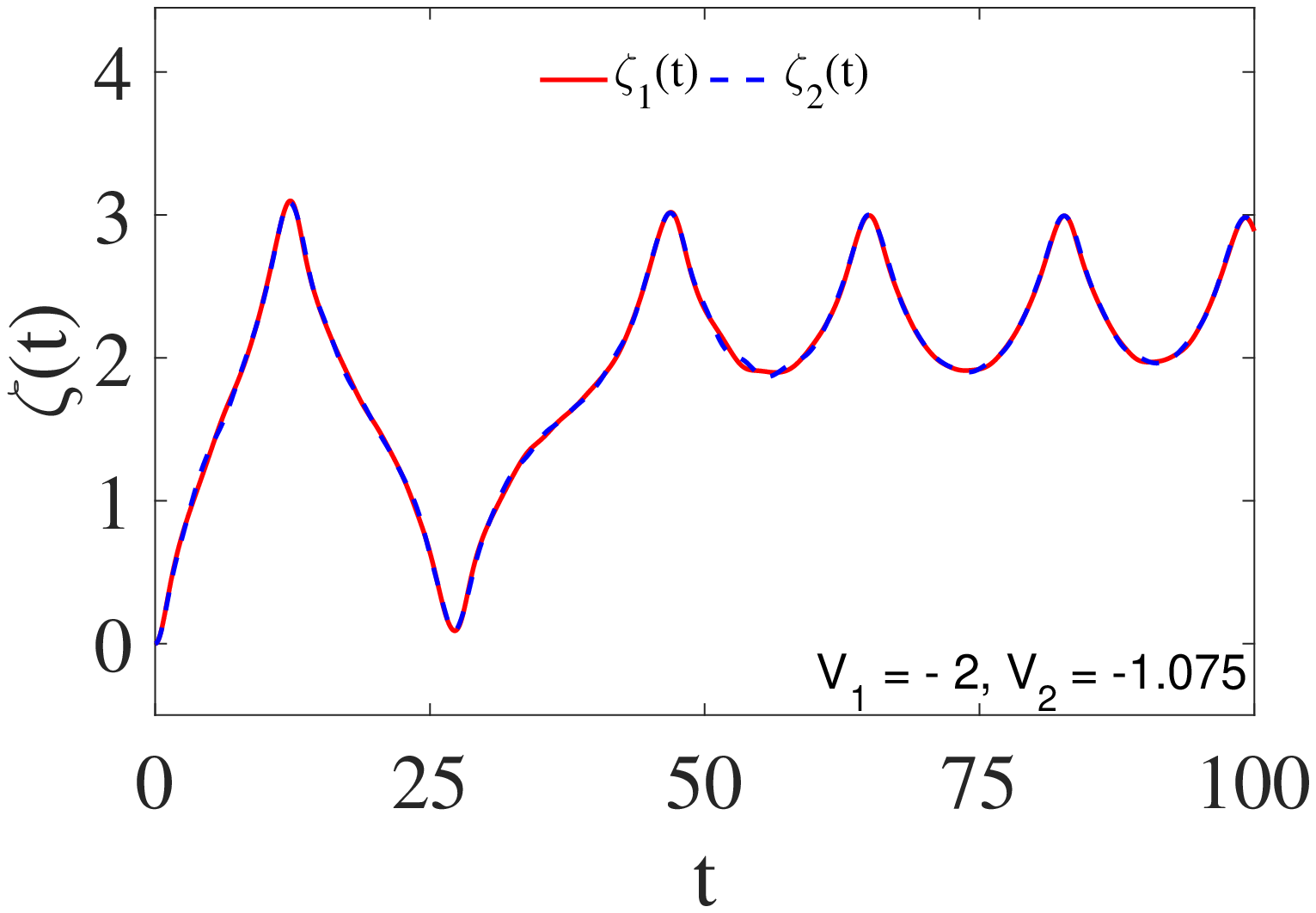}
\hskip -.45cm
\includegraphics[scale=0.38]{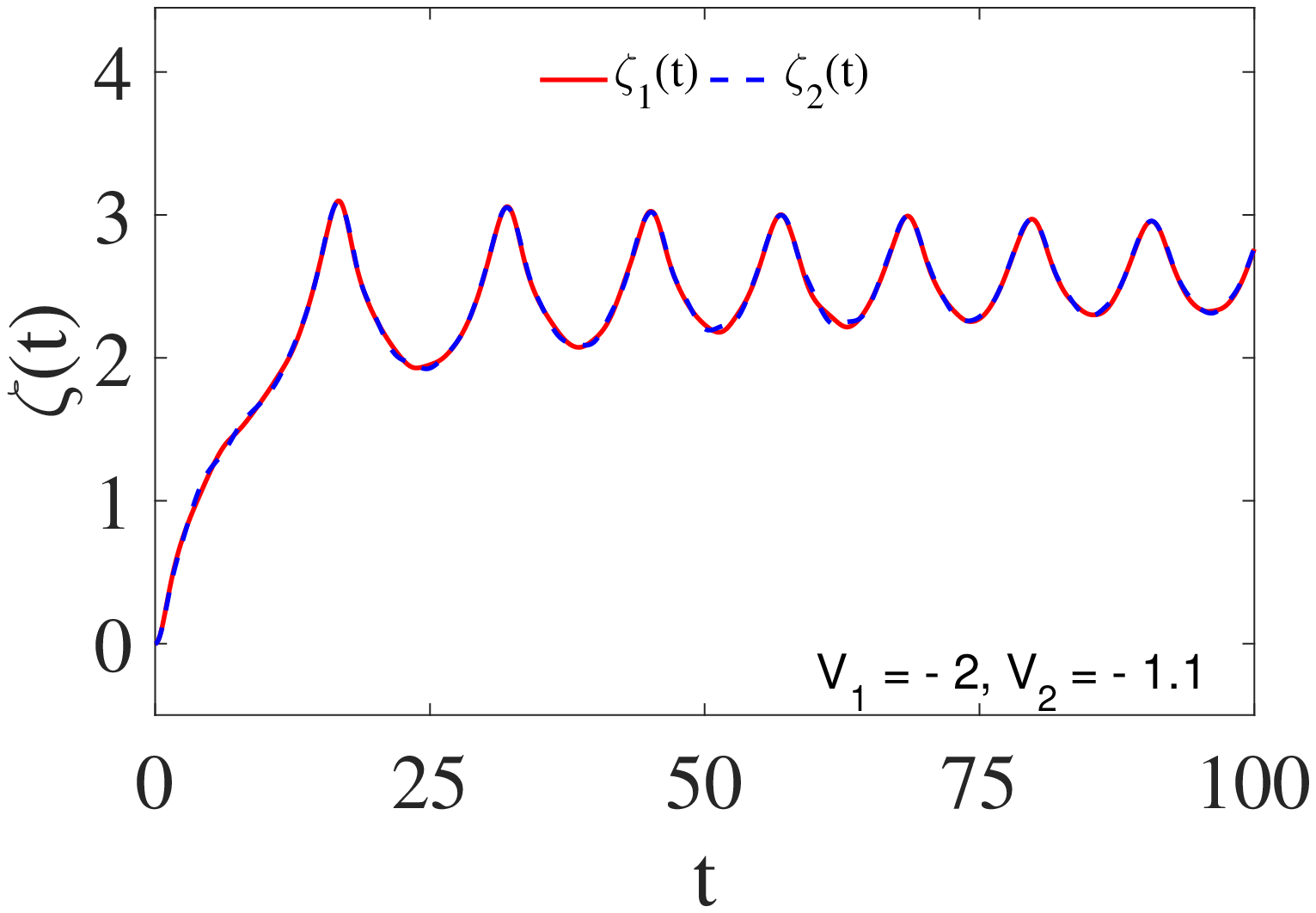}}
\caption{\small The center of mass coordinate of the first (red solid) and the second (blue dashed) components 
of binary BEC in a parabolic trap $\beta(y-y_0)^2$ for different $V_1, V_2$ values indicated in the panels. 
Other parameter are fixed as:  $\gamma_{12}=-2$, $\gamma_{1}=-2.0$, $\gamma_2=0$, ${\cal N}_1=3.5$, ${\cal 
N}_2=2.5$, $\beta=0.1, $ $y_0=0.5\pi$.
}
\label{fig9}
\end{figure*}
\begin{figure}
\centerline{
\includegraphics[clip=true,scale=0.55]{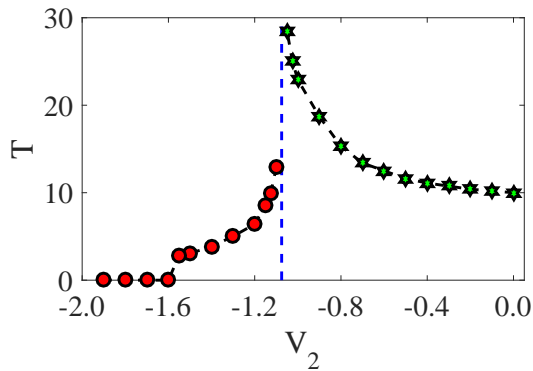}}
\caption{\small Period T  of the binary soliton  versus $V_2$ for $0<|V_2|<|{V_2}_{cr}|$ (green stars) and for 
$|V_2|>|{V_2}_{cr}|$ (red circles). Parameters are fixed as in Fig. \ref{fig9} for which  ${V_2}_{cr}\approx -1.075$.}
\label{fig10}
\end{figure}
\begin{figure*}
\centerline{
\hskip -.2cm
\includegraphics[clip=true,scale=0.44]{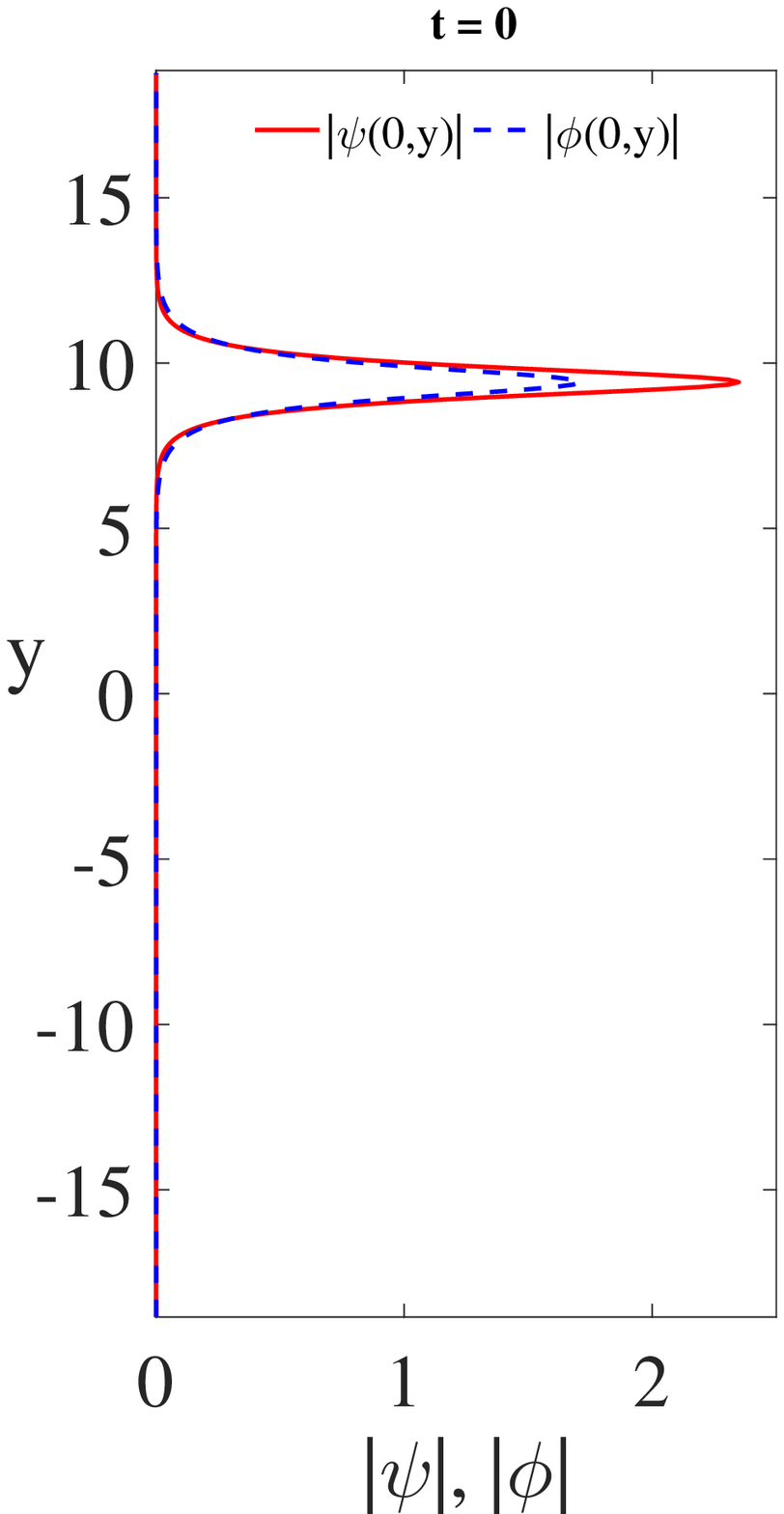}
\hskip -1.cm
\includegraphics[ clip=true,scale=0.44]{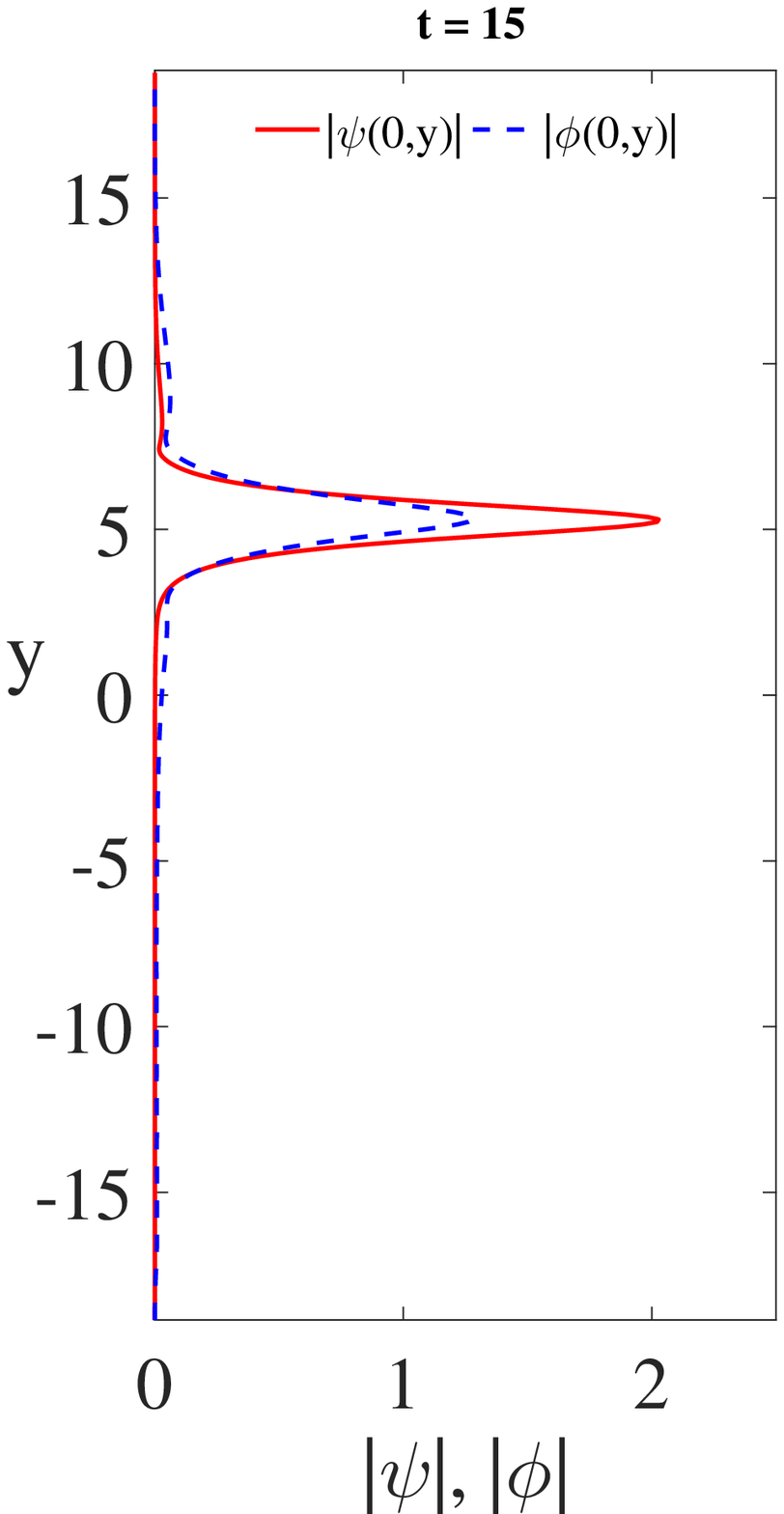}
\hskip -1.cm
\includegraphics[clip=true,scale=0.44]{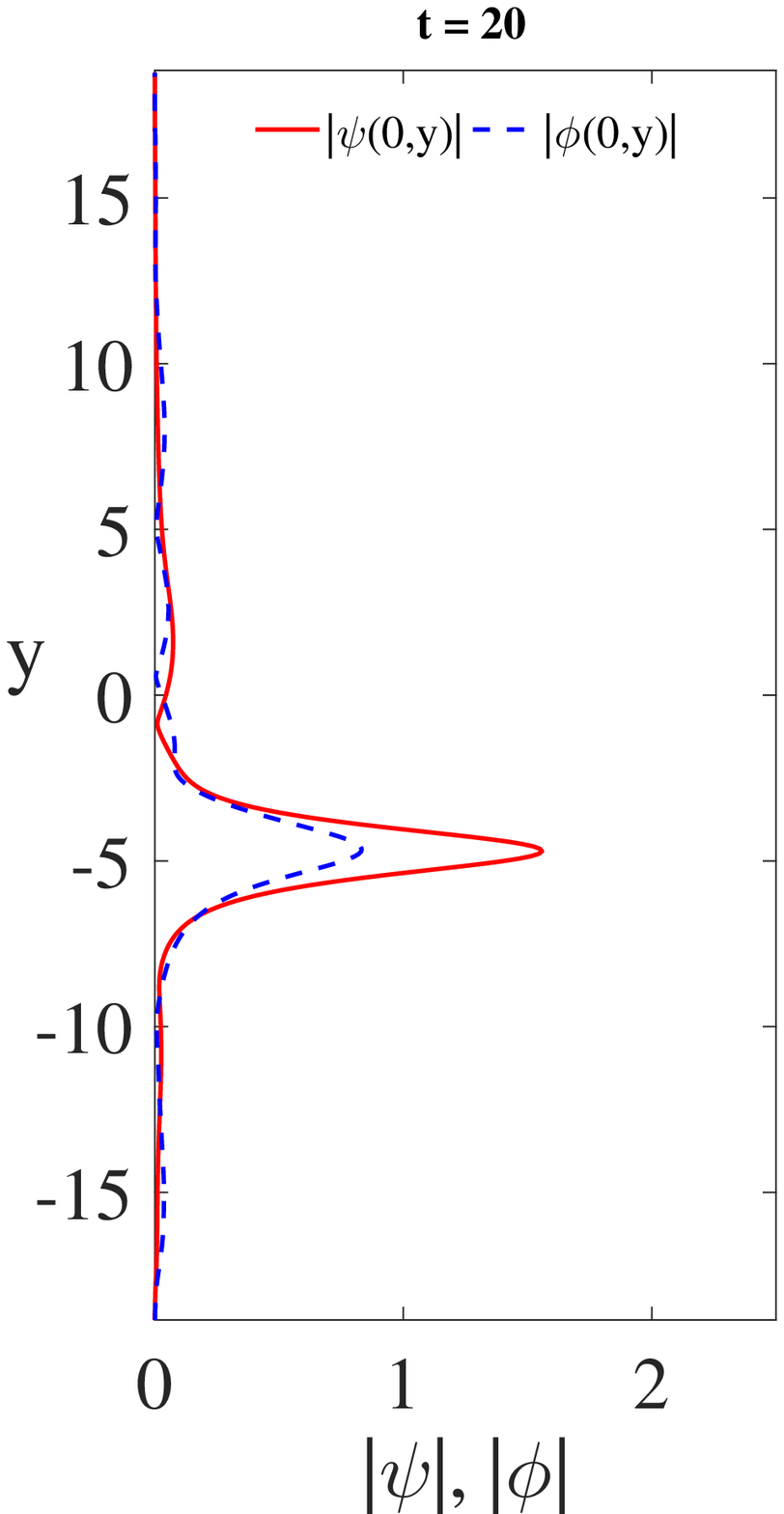}
\hskip -1.cm
\includegraphics[ clip=true,scale=0.44]{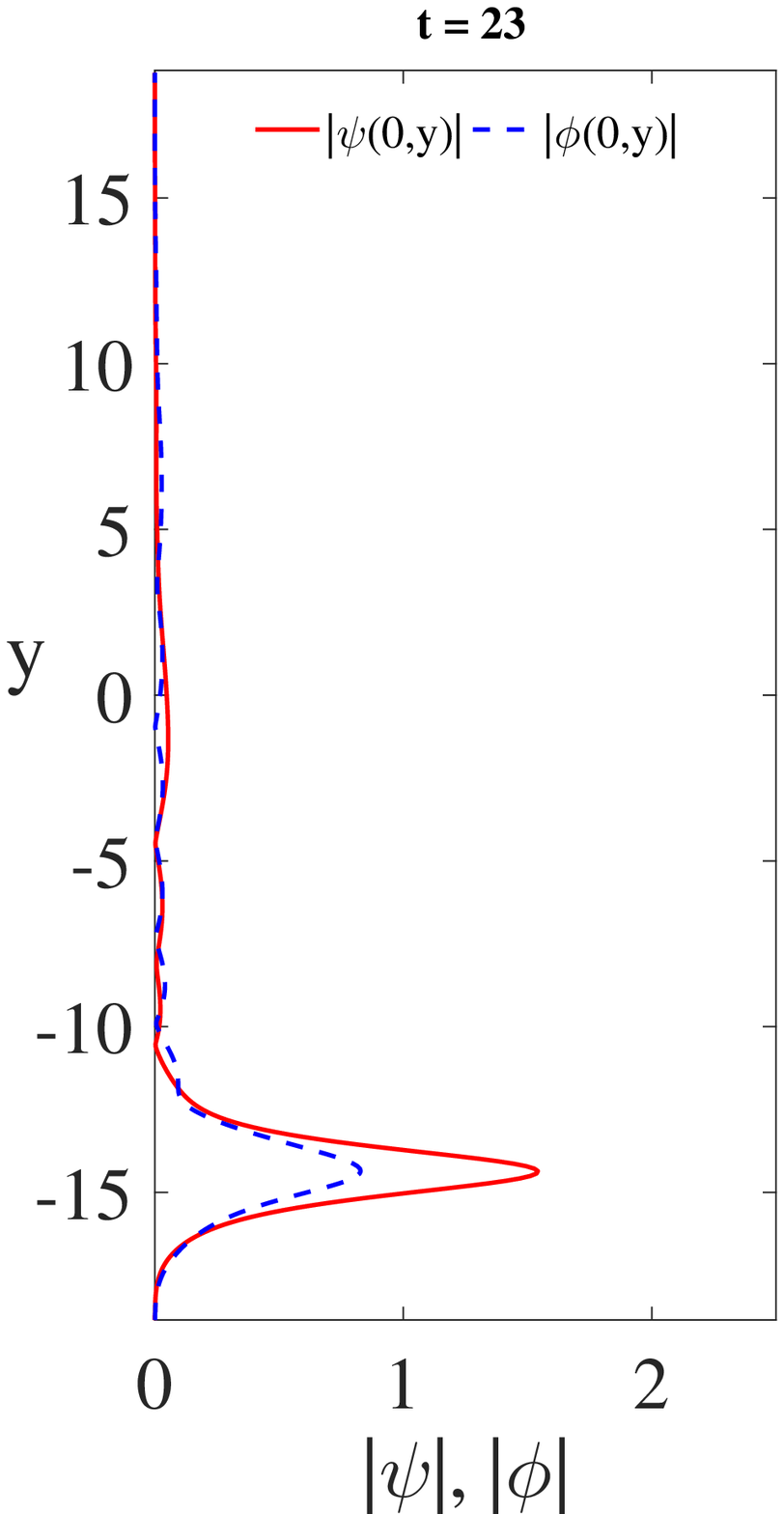}}
\caption{\small Density sections at $x=0$ of the two components $|\psi(0,y)|$ (continuous red lines) and $|\phi(0,y)|$ (dashed blue 
lines) at different instant of times, $t=0, 15, 20, 23$, starting from the left, respectively. The parameter $\alpha=0.155$ is just 
slightly above the critical value $\alpha_{cr}=0.15$. Other parameters are fixed as:  $\gamma_{12}=-2.0$, $\gamma_{1}=-2.0$, 
$\gamma_2=0$, $V_1=-2.0$, $V_2=-1$, ${\cal N}_1=3.5$, ${\cal N}_2=2.5$.}
\label{fig11}
\end{figure*}
\begin{figure}
\centerline{
\includegraphics[clip=true,scale=0.5]{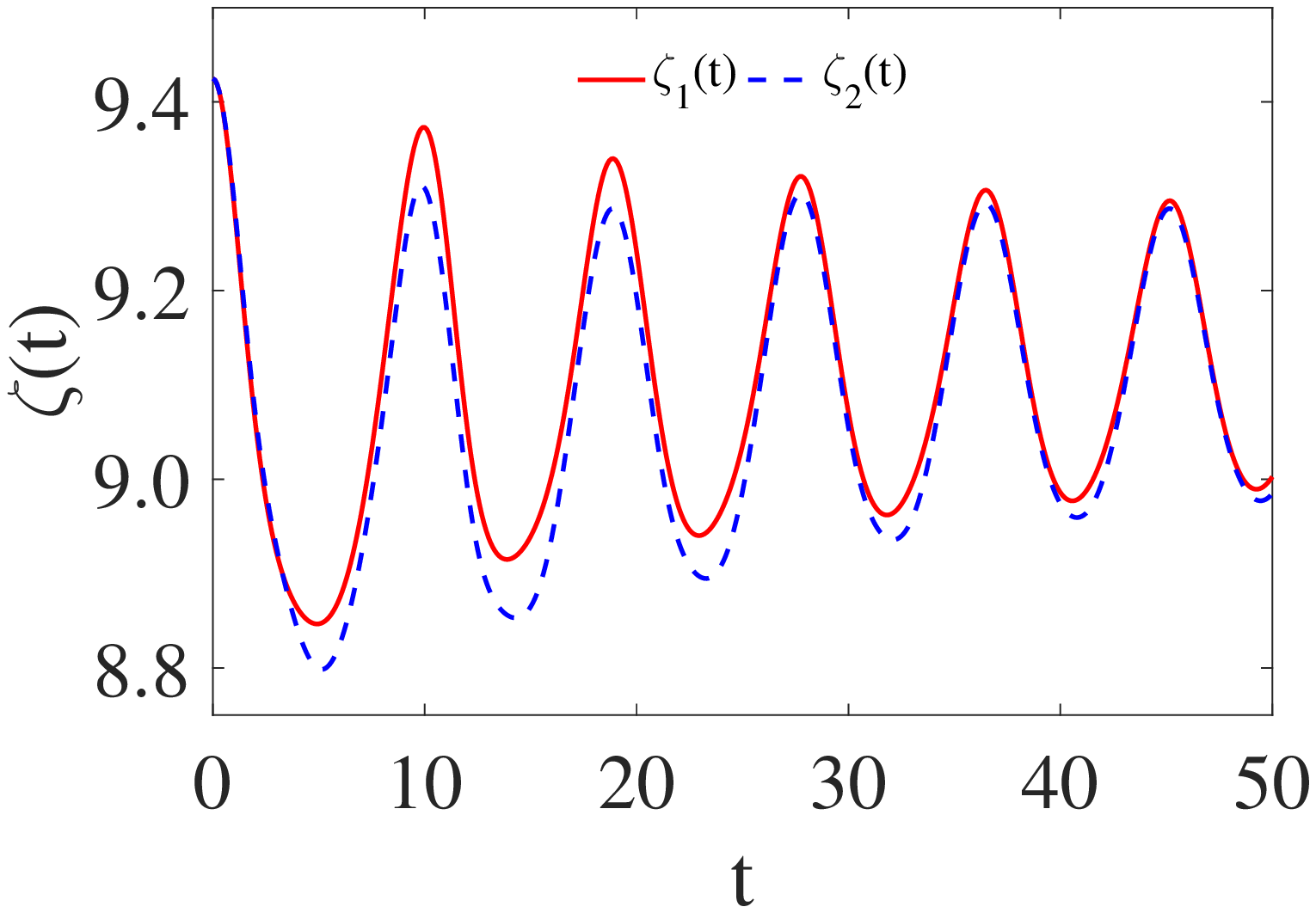}}
\caption{\small Time evolution of the COM coordinates of the first (red solid) and the second (blue dashed)  
components of a binary BEC in the linear ramp potential $\alpha(y-y_0)$ with $\alpha=\alpha_{cr}=0.15$. Other 
parameters are fixed as: $\gamma_{12}=-2.0$, $\gamma_{1}=-2.0$, $\gamma_2=0$, $V_1=-2.0$, $V_2=-1$, ${\cal N}_1=3.5$, 
${\cal N}_2=2.5$, $y_0=3\pi$. Notice that this parameter setting is the same as for the point 
$\alpha_{cr}=0.15$ in Fig.~\ref{fig14} }
\label{fig12}
\end{figure}
\subsection{Parabolic trap}
A parabolic trap of the form  $\beta (y-y_0)^2$, with $\beta$ and $y_0$ real parameters controlling the strength 
and the position of the trap minimum respectively, is  applied to both  components of the binary matter wave. 
The initial condition is taken as  a stationary binary soliton  at the position $x=0, y=0$ obtained from 
imaginary time with the trap minimum fixed at $y_0=0$. In order to put the soliton in motion we shift  the trap 
minimum from $y_0=0$  to a nonzero value at $t=0$ along the $y-$axis and compute the resulting dynamics from 
numerical time integration of the GPE  system.

In Fig.~\ref{fig7} we show the dynamics of the COM of the two BEC components for different values of the 
initial shift $y_0$ in absence of the NOL (i.e with only the LOL acting is in the $x-$direction). We see that in 
this case the motion of the two  components is perfectly harmonic with the oscillation period independent of the 
amplitude. In the presence of the NOL, however, the situation drastically changes.

As it is seen from  Fig.~\ref{fig8}, when the strength of the NOL is small the binary BEC executes oscillations around the minimum 
of the harmonic trap, as expected, but the frequency of the oscillation depends on the amplitude, i.e. the presence of the NOL 
introduces anharmonicity. By increasing the strength of the NOL, while keeping all other parameters fixed, a 
critical value of $V_2$ appears (i.e. ${V_2}_{cr} \approx -1.075$ in Fig.~\ref{fig9})  above which the self-
trapping phenomenon occurs. Just before the self-trapping transition the dynamics strongly  deviates from 
harmonic motion with the trajectory of the center of mass of the binary soliton acquiring zig-zag like profiles. 

Note that the oscillation below and above the critical point have different nature, the first   occurs around 
the minimum of the parabolic trap (fixed to $\pi/2$ in the figure), the second around the minimum of the 
effective NOL potential (evaluated at the self-trapping position). In both cases, however,   the two components 
oscillate together on a long time scale with their centers of mass practically overlapped. 
As the  strength of the NOL is further increased beyond the transition point the soliton becomes self-trapped 
and the oscillations occur inside the NOL effective potential. In Fig.~\ref{fig10} we show the dependence on 
$V_2$ of the period, T,  of the binary soliton when the oscillation occurs inside the parabolic trap and inside the 
effective NOL potential, i.e. for $0<|V_2|<|{V_2}_{cr}|$ and for $|V_2|>|{V_2}_{cr}|$, respectively. 
The above dynamical  behaviors can be  qualitatively understood in terms of the NOL potential 
barrier discussed in the previous section. When the NOL strength is small compared to the critical value the 
soliton motion can easily overcome the   barrier and the resulting oscillation is harmonic but with smaller 
frequency with respect to the case $V_2=0$. The larger oscillation period observed for  $|{V_2}_{cr}| \le 
|V_2| \ge 0$ can be ascribed to the slowing down of the soliton during the  overcoming of the barrier. This slowing 
down effect  increases as $V_2$ approaches the critical value at which the binary soliton  becomes self-trapped. 
The oscillations in the self-trapped regime $|V_2|>|{V_2}_{cr}|$ obviously depend on the NOL effective potential at 
the position where the self-trapping occurs. 
As $V_2$ is increased above the critical value amplitudes and periods of these oscillations decrease, as expected for states 
localized in strong trapping potentials. For very large NOL amplitudes the binary soliton becomes fully rest at a position 
that, in the absence of the NOL,  would be of non equilibrium for a  parabolic trap. 

\subsection{Linear ramp potential}
\begin{figure}
\centerline{
\includegraphics[trim=0cm 0cm 0cm 0cm, clip=true,scale=0.7]{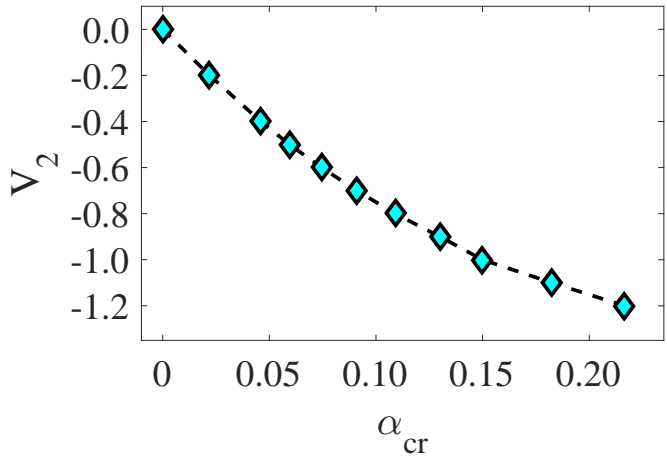}}
\caption{\small Critical curve in the parameter plane $V2, \alpha$ separating the falling regime (region above 
the curve) from the dynamical self-trapping regime  (region below the curve). Other parameters are fixed as: 
$\gamma_{1}=-2.0$, $\gamma_2=0$, $\gamma_{12}=-2$, $V_1=-2.0$, ${\cal N}_1=3.5$, ${\cal N}_2=2.5$. The dependence of $\alpha_{cr}$ 
to the inter component interaction coefficient $\gamma_{12}$. For 
numerical purpose we fixed values of different parameters at $\gamma_{1}=-2.0$, $\gamma_2=0$, $V_1=-2.0$, 
$V_2=-1$, ${\cal N}_1=3.5$, ${\cal N}_2=2.5$.}
\label{fig13}
\end{figure}

\begin{figure}
\centerline{\includegraphics[trim=0cm 0cm 0cm 0cm, clip=true,scale=0.7]{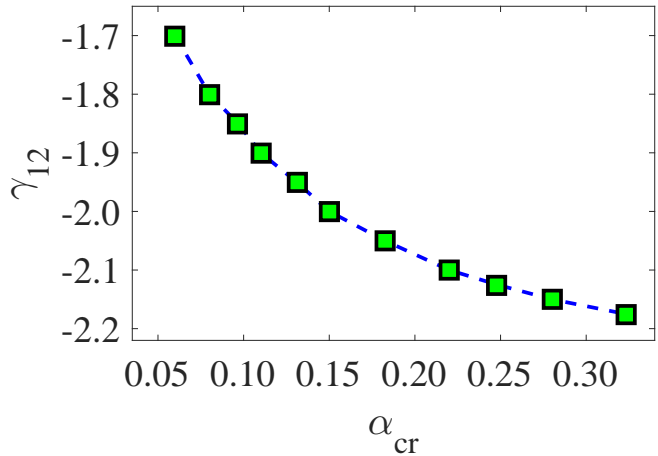}}
\caption{\small Critical curve in the parameter plane $\gamma_{12}, \alpha$ separating the falling regime (region 
above the curve) from the dynamical self-trapping regime (region below the curve). Other parameters are fixed as: 
$\gamma_{1}=-2.0$, $\gamma_2=0$, $V_1=-2.0$,  $V_2=-1.0$, ${\cal N}_1=3.5$, ${\cal N}_2=2.5$. 
}
\label{fig14}
\end{figure}

In this section we  consider a linear ramp potential of the form $\alpha\,(y-y_0)$,  with $\alpha, y_0$ real parameters,  mimicking 
a gravitational field acting on both components in the $y-$direction.
The initial conditions are  taken as stationary solutions of the  $\alpha=0$ case located at the position  $x_0=0, y_0=3\pi$. The 
linear ramp potential  is switched on at $t=0$, and the time evolution is obtained from direct numerical integrations of the GPE. 
Numerical results  are displayed in Figs.~\ref{fig11}-\ref{fig14}.

From Fig.~\ref{fig11} we see that for sufficiently small values of the strength of the NOL and for $V_2=0$ (not 
shown for brevity), the binary matter wave is  accelerated just as  expected for solitons falling in a 
gravitational field, with only some small deformations of the profiles and some  tiny emission  of radiation 
that  are probably due to  the sudden acceleration and switching-on of the potential.  By increasing the 
strength of the NOL and keeping all other parameter fixed, however,  we find that there  exist a critical value 
of $V_2$ above which, quite remarkably, the binary soliton instead of falling under the action of the gravity  
remains suspended, executing oscillations around a fixed position. This is the same DSTP discussed for the parabolic case before. 

The critical thresholds for the falling or the DSTP of the binary soliton depend on all system parameters and 
particularly on the slope of the ramp, the strength of the NOL, and the inter-component interaction. In Figs.~\ref{fig13},~
\ref{fig14} we show curves in the parameter planes $(V_2, \alpha)$ and $(\gamma_{12}, \alpha)$ that separate the falling regime 
(region above the curves) from the dynamical self-trapping regime (region below the curves), for a specific set of the rest of 
parameters, respectively. 

From these figures we  see that a stronger gravitational field (i.e. larger slope of the ramp) requires a 
stronger NOL amplitude $V_2$ or a larger inter-component interaction $\gamma_{12}$,  for the dynamical self-
trapping phenomenon to occur. These behaviors  can be easily understood  in terms of the effective self-trapping 
barrier. Indeed, an increase of the ramp's slope implies a larger soliton energy which in turn requires a 
larger potential barrier to stop it. 

From Eq.~(\ref{eff-pot}) it is clear  that the effective NOL potential can 
be increased either by increasing $V_2$ (this explaining the curve in Fig.~\ref{fig13}), or by increasing the 
density of the soliton which can be achieved by increasing the attractive inter-component interaction so that 
the the binary soliton becomes more focused (this explains the curve in Fig.~\ref{fig14}).  

In closing this section we  remark that in principle one  could apply the external potentials  in the $x$ 
direction instead of the $y-$direction. We discard this possibility, however, for the following two reasons. 
First, in this setting  there would be no dynamics in the $y-$direction and therefore no DSTP would occur. 
Second, in the ramp potential case the combined action with the LOL would lead, in 
analogy to what happens in the 1D single component case (see \cite{salerno}), to  dynamical instabilities that 
would destroy the binary soliton.   

\section{Discussion and Conclusions}

Before closing this paper we feel compelled to discuss advantages of  2D cross-combined LOL and NOL settings and  
possible practical implications  of our results. In general, for the development of  soliton applications it is important:  i) 
their stability,   ii) their management i.e. the possibility to manipulate their motion.  

In a multi-dimensional setting  the first point is already non trivial due to presence of collapse, delocalization, etc. It is 
possible to avoid these adverse phenomena, at least in a region  of nonzero measure in the parameter space, by exposing the 
condensate to the action of a 1D LOL (a 2D LOL would also  stabilize the solitons against collapse but it would limit their 
mobility). This is true both for the single component case, as demonstrated in \cite{salerno1}, and for binary solitons,  
as one can see from the  $V_2=0$ cases of Figs. \ref{fig5}, \ref{fig8}. The  presence of a NOL in the $y-$direction, although not 
strictly necessary for existence and stability, allows to satisfy the requirement ii).  Indeed, the control of the DSTP can be 
used as a tool for moving or stopping solitons in given positions as shown in the examples discussed above. It is remarkable that 
in our setting the management can be done without any physical modification of the system, simply by acting on the external laser 
fields that control the interactions via the usual (two-body) or the optically induced (NOL) Feshbach resonances. The management of 
the soliton motion is certainly a fundamental step for experimental and applicative developments.

In conclusion we have demonstrated, both by variational analysis and by direct numerical integration of the GPE 
coupled equations,  the existence and stability of  2D binary BEC mixtures trapped in a cross-combined lattice 
consisting of a one-dimensional linear optical lattice in the x-direction for the first component and a 1D non-
linear OL in the $y$-direction for the second component.
%%%%%%%%%%%%%%%%%%%%%%%%%%%%%%%%%%%%%%%%%%%%%%%%%%%%
Dynamical properties of such  binary 2D soliton have been investigated both by phase imprinting and by applying 
additional external potentials along the constraint direction of the cross combined OLs. In particular, we have  shown the 
occurrence of the DSTP that allows to hold  a soliton at rest in a non equilibrium position of a parabolic potential, 
or to prevent a soliton from falling under the action of gravity. The existence of thresholds in the parameter space for the 
occurrence of these phenomena have been also demonstrated.
%%%%%%%%%%%%%%%%%%%%%%%%%%%%%%%%%%%%%%%%%%%%%%%%%%%%%%%%%%%%%%%%%%%%%%%%%%%%%

\section*{Acknowledgments}
  KKI acknowledges  financial support from the Ministry of Innovative Development of the Republic of Uzbekistan  for a three months 
  grant under the "Short Term Scientific Internship of Young Scientists in Foreign Scientific Organizations" program, contract No. 
  74 with the University of Salerno. KKI also acknowledges the  Physics Department E.R. "Caianiello" for the hospitality and for 
  the internship opportunity during which this work was completed. The authors wish to thank Dr. B.B. Baizakov for a critical 
  reading of the manuscript.


\begin{thebibliography}{99}
\bibitem{note3} We find that in the case $V_2=0$  the two trial functions (sech-sech and  Gaussian),  lead to similar  variational 
results but the  comparison with the exact numerical results in the top panels of Fig.\ref{fig3} is better for the Gaussian ansatz.
\bibitem{note} For non cross-combined lattices, e.g. a 2D LOL or a 2D NOL, the two minimizations would be compatible.
\bibitem{note2} The phenomenon of dynamic self-trapping exists also in 1D environments for ordinary (i.e. single-component) BEC 
solitons. This was apparently unnoticed  in previous studies.
\end{thebibliography}
\end{document}